\documentclass[aps,pra,floats,floatfix,twocolumn]{revtex4}

\usepackage{ifthen}
\usepackage{ifpdf}
\usepackage{color}

\ifpdf
\usepackage{graphicx}
\usepackage{epstopdf}
\else
\usepackage{graphicx}
\usepackage{epsfig}
\fi
\graphicspath{{./FigsN/}{./Figs/}{./}}

\usepackage{latexsym}
\usepackage{amsmath}
\usepackage{amssymb}
\usepackage{bm}
\usepackage{wasysym}

\usepackage{hyperref}

\newcommand{\const}{\mbox{const}}
\newcommand{\trc}{\mbox{trace}}

\newcommand{\eexp}{\mbox{e}^}

\newcommand{\mass}{\mathsf{m}}

\newcommand{\tbox}[1]{\mbox{\tiny #1}}
 
\newcommand{\mylabel}[1]{\label{#1}} 
\newcommand{\beq}{\begin{eqnarray}}
\newcommand{\eeq}{\end{eqnarray}} 
\newcommand{\be}[1]{\begin{eqnarray}\ifthenelse{#1=-1}
{\nonumber}{\ifthenelse{#1=0}{}{\mylabel{e#1}}}}
\newcommand{\ee}{\end{eqnarray}} 

\newcommand{\Eq}[1]{\textcolor{blue}{Eq.\!\!~(\ref{#1})}} 
\newcommand{\Fig}[1]{\textcolor{blue}{Fig.}\!\!~\ref{#1}} 
\newcommand{\hide}[1]{}
\newcommand{\rmrk}[1] {{#1}}  
\renewcommand{\cite}[1]{\textcolor{blue}{[\onlinecite{#1}}]} 
\newcommand{\hrefl}[1]{}

\begin{document}

\title{Triangular Bose-Hubbard trimer as a minimal model for a superfluid circuit}

\author{Geva Arwas$^1$, Amichay Vardi$^2$, Doron Cohen$^1$}

\affiliation{
\mbox{$^1$Department of Physics, Ben-Gurion University of the Negev, Beer-Sheva, Israel} \\
\mbox{$^2$Department of Chemistry, Ben-Gurion University of the Negev, Beer-Sheva, Israel}
}

\begin{abstract}
The triangular Bose-Hubbard trimer is topologically the minimal model for a BEC superfluid 
circuit. As a dynamical system of  two coupled freedoms it has mixed phase-space with chaotic dynamics. We employ a semiclassical perspective to study triangular trimer physics beyond the conventional picture of the superfluid-to-insulator transition. From the analysis of the Peierls-Nabarro energy landscape, we deduce the various regimes in the $(\Omega,u)$ parameter-space, 
where~$u$ is the interaction, and $\Omega$ is the superfluid rotation-velocity. We thus characterize the superfluid-stability and chaoticity of the  many-body eigenstates throughout the Hilbert space.  
\end{abstract}

\maketitle


\section{Introduction}

The experimental study of Bose-Einstein Condensates (BECs) 
allows the realization of ultracold atomic superfluid circuits \cite{e1,e2,e3,e4,e5,e6,e7,e8,e9,e10,e11,e12,e13,e14} 
and the incorporation of a laser-induced weak-link (a bosonic Josephson junction) within them \cite{e15}. 
By rotating such a barrier it is possible to {induce current \cite{Hekk} } and to drive phase-slips 
between quantized superfluid states of a low dimensional toroidal ring \cite{e15}.  
Such mesoscopic devices open a new arena for detailed study of complex Hamiltonian dynamics.  

\begin{figure}[b!]
\includegraphics[width=8.5cm]{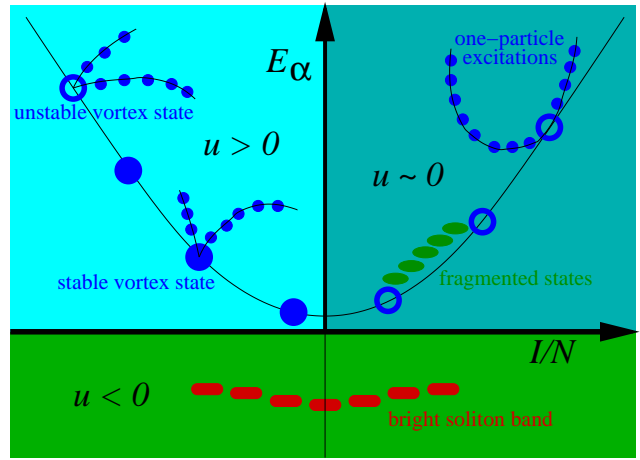}

\caption{(Color online)
Representative eigen-energies $E_{\alpha}$ of a rotating 
ring system are classified by the current $I$ they carry 
in the rotating frame of reference. 
The filled and hollow blue circles represent stable and unstable 
vortex-states for which $I/N$ is quantized.  
The blue dots represent their Bogolyubov one-particle excitations. 
On the upper right $u$ is vanishingly small, 
while on the upper left $u$ is large and positive, resulting in phonon branches.
The green elliptic dots represent fragmented states 
(here the particles are divided between two vortex orbitals). 
For negative $u$ (lower plane) there are stable bright 
soliton solutions. Red oval dots represent eigenstates
that are formed by superposition of such solitons.
For positive $u$, and appropriate rotation frequency,  
one may observe additional type of eigenstates 
that are formed of dark solitons (not illustrated).
Adding weak disorder, the unstable vortex-states 
would mix with their excitations, 
and the soliton band would decompose into 
a set of standing solitons.   
} 

\label{f1}
\end{figure}

The hallmark of superfluidity is a stable non-equilibrium steady-state current. 
If $N$ bosons in a rotating ring are condensed into a single plane-wave orbital, 
one obtains a ``vortex state" with a quantized current per particle ($I/N$) \cite{Udea,Carr1,Carr2,Brand1}. 
For non-interacting bosons the lowest vortex-state is also the ground-state. 
It is stable and carries a microscopically small ``persistent current". 
By contrast, all higher vortex states are unstable. 
Interactions change the picture dramatically \cite{nir}: 
the Bogolyubov-spectrum of the one-particle excitations of a vortex-state is modified 
(e.g. by the appearance of phonons), 
and hence all vortex-states that satisfy the Landau criterion \cite{Landau,Hakim,Leboeuf} become stable. 
See \Fig{f1} for illustration, and {Section~VI} for extra pedagogical details.

The vortex-state of bosons in a ring constitutes one particular example 
of a coherent (non-fragmented) state. Other coherent-state solutions 
may correspond, for example, to all bosons condensed in a localized orbital (bright soliton)~\cite{Udea}, 
or in an orbital that has a notch (dark soliton) \cite{Carr1,Carr2}. 
From such non-stationary classical solutions one can superpose stationary quantum eigenstates whose 
angular momentum is not quantized.        

The above picture is missing a central ingredient: there is no reference 
to the global structure of the underlying phase-space that dictates the dynamics. 
Vortex-states and solitons are minima or maxima of the energy landscape, 
and the Bogolyubov spectrum merely reflects the linear-stability analysis in 
the vicinity of these solutions.  We are therefore motivated to consider the simplest paradigm for a superfluid circuit which still allows the thorough investigation of its phase-space structure. 
The natural choice for such a model is bosons in a one-dimensional 
ring as in Ref.~\cite{Brand1} or its discrete $M$ site version, described 
by the Bose-Hubbard Hamiltonian (BHH) \cite{BHH1,BHH2,Ofir1,Ofir2} as in Ref.~\cite{Altman}. 

The phase-space of the ring model was studied within a two-orbital approximation \cite{Brand2}. However,  such an approximation is not a valid minimal model by itself. From a {\em topological} point of view the minimal model for a superfluid circuit has to involve $M{=}3$ sites: a triangular Bose-Hubbard trimer (this would be equivalent to three non-localized modes).  
A close relative is the linear Bose-Hubbard trimer. The trimer phase-space has been partially studied in several papers 
\cite{trimer1,trimer2,trimer3,trimer4,trimer5,trimer6,trimer7,trimer8,trimer9,trimer10,trimer11,trimer12,trimer13}, 
and has been recognized as a building block for studies of 
transport \cite{Henn1,Henn2} and mesoscopic thermalization \cite{KottosBoris,trm}.

In this work we study the triangular Bose-Hubbard trimer as a topologically minimal model for a BEC superfluid circuit. In particular, we note that the mixed phase-space aspect of the dynamics has far reaching consequences, producing non-trivial physics that goes beyond the conventional picture of the Mott superfluid-to-insulator transition. Our approach relies on the analysis of the Peierls-Nabarro energy landscape \cite{Henn1,Henn2,Kivshar93,Rumpf04}, from which we deduce various regimes in the $(\Omega,u)$ parameter-space of the triangular BHH trimer,    
where~$u$ is the interaction parameter, and $\Omega$ is the superfluid rotation-velocity.
In each of these regimes we outline the structure of phase-space and the classification of the many-body eigenstates. The criteria for the superfluid stability as opposed to chaoticity are thus determined.

\begin{figure*}[t!]

(a) \hspace{6cm} (b) \\
\includegraphics[height=5cm]{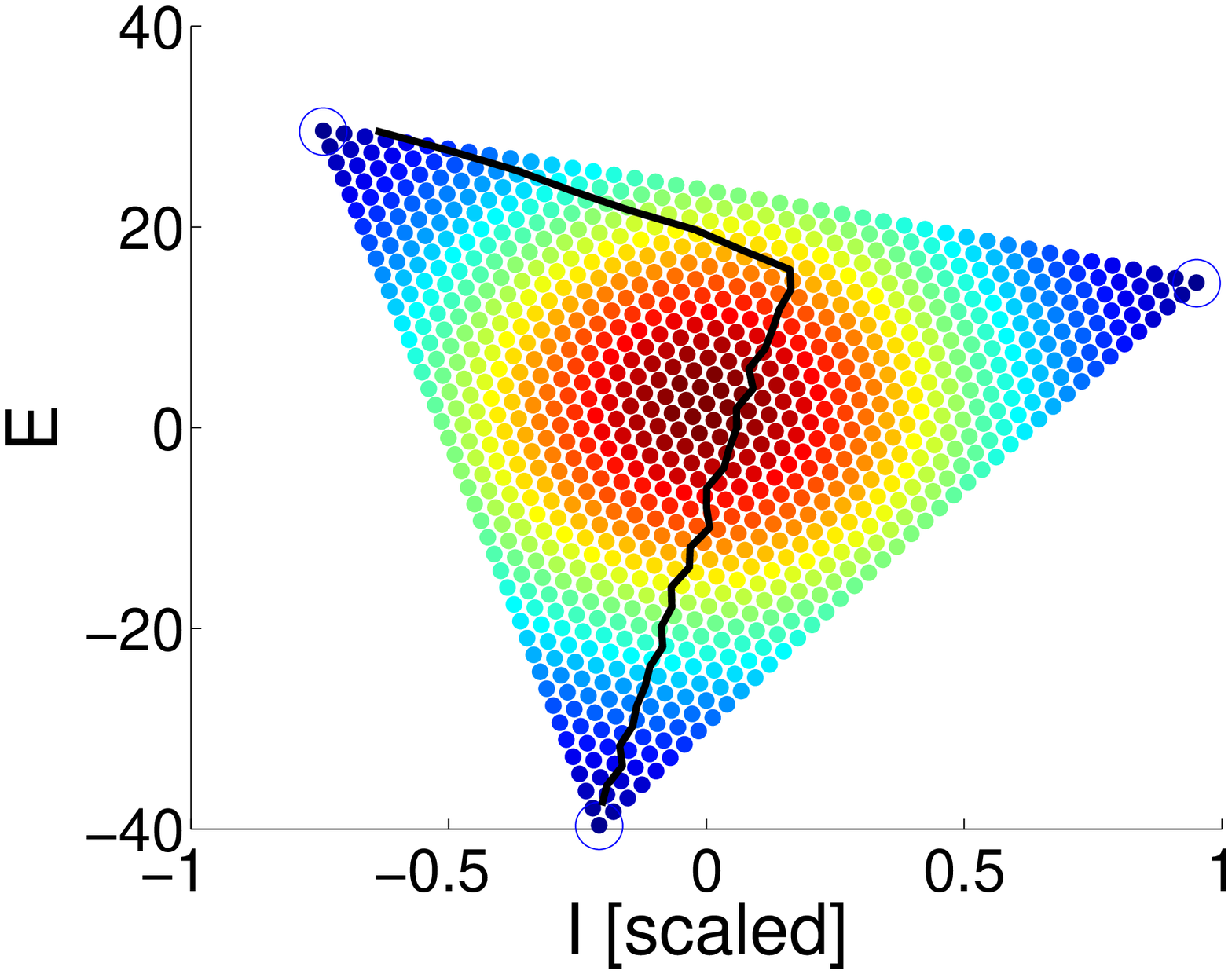}
\includegraphics[height=5cm]{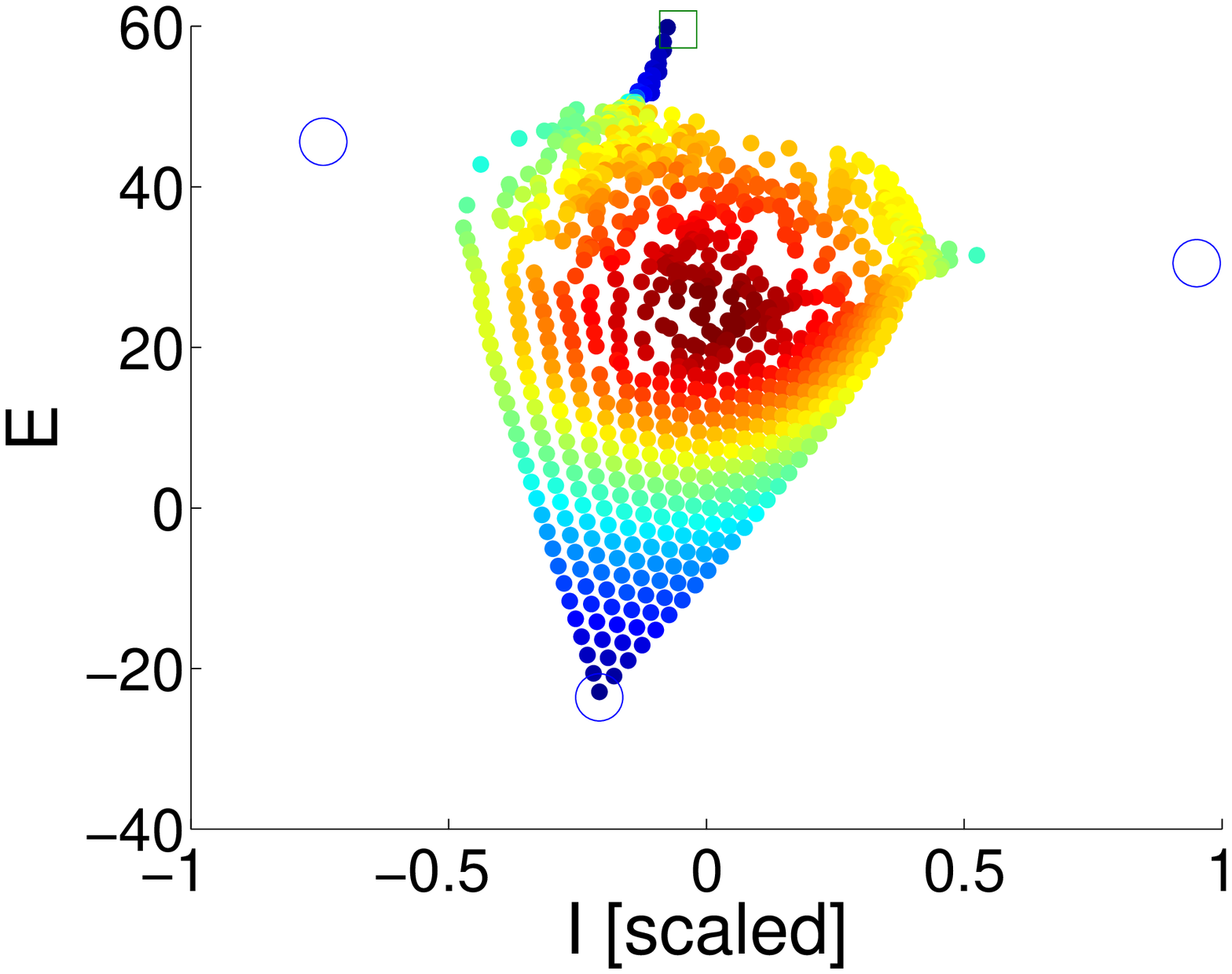}
\hspace{2.2cm}
\includegraphics[height=5cm]{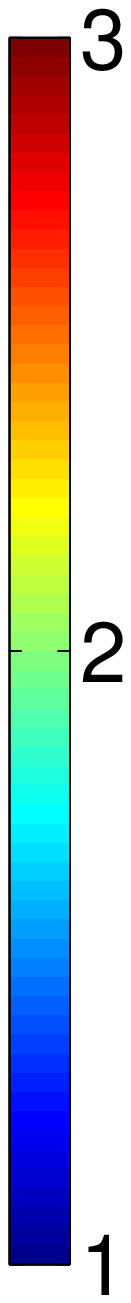}

\ \\

(c) \hspace{6cm} (d) \hspace{5cm} (e) \\
\includegraphics[height=5cm]{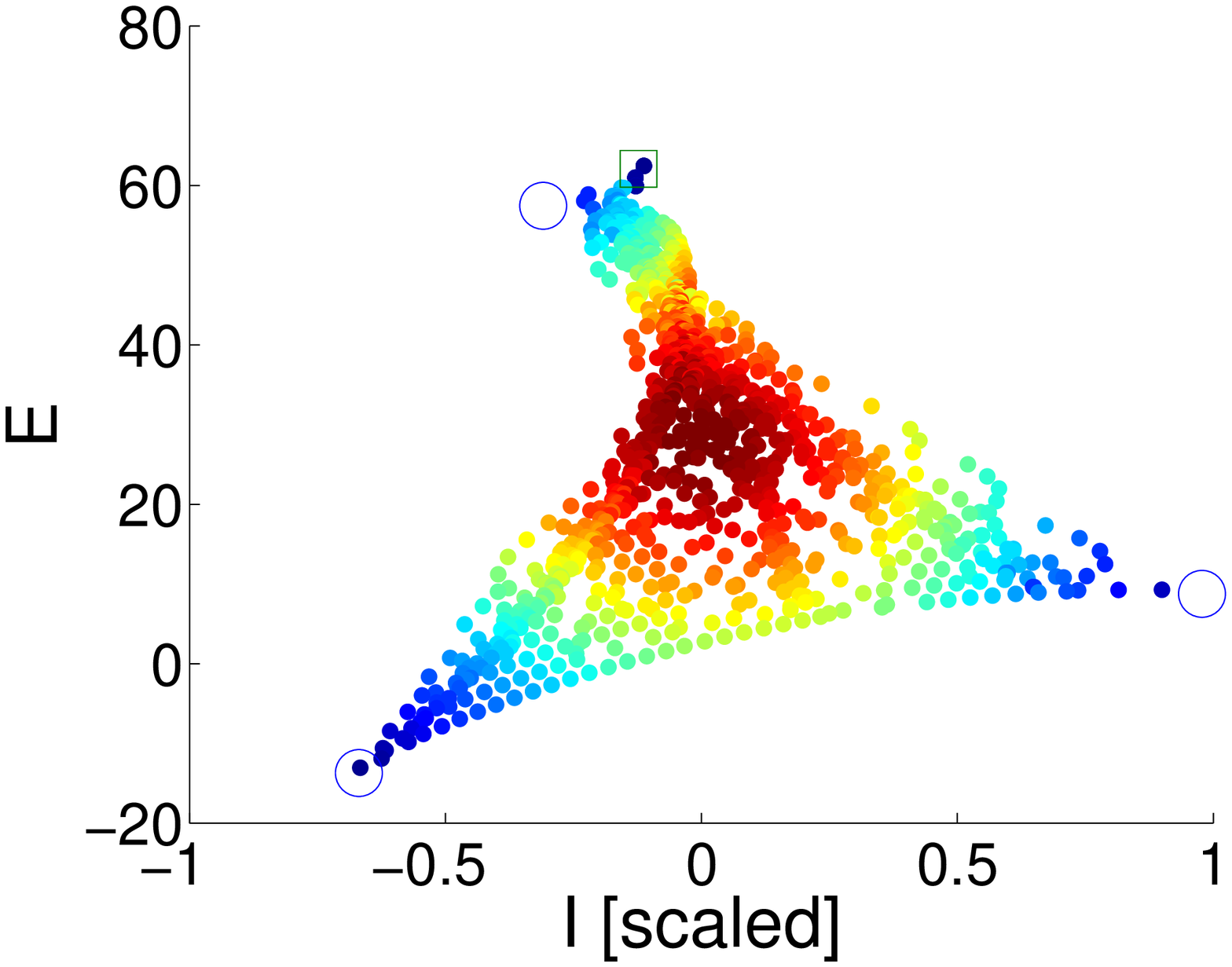}
\includegraphics[height=5cm]{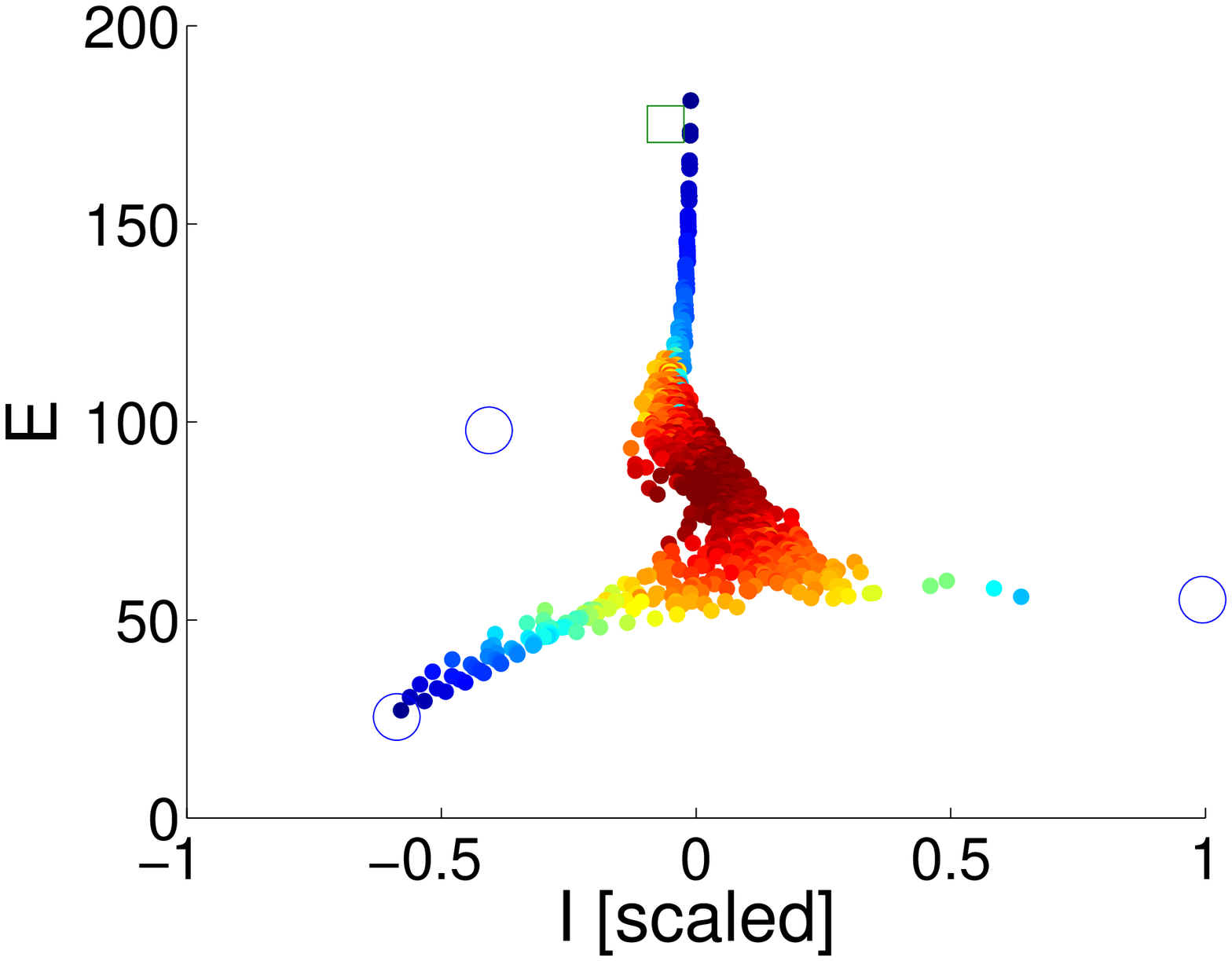}
\includegraphics[height=5cm]{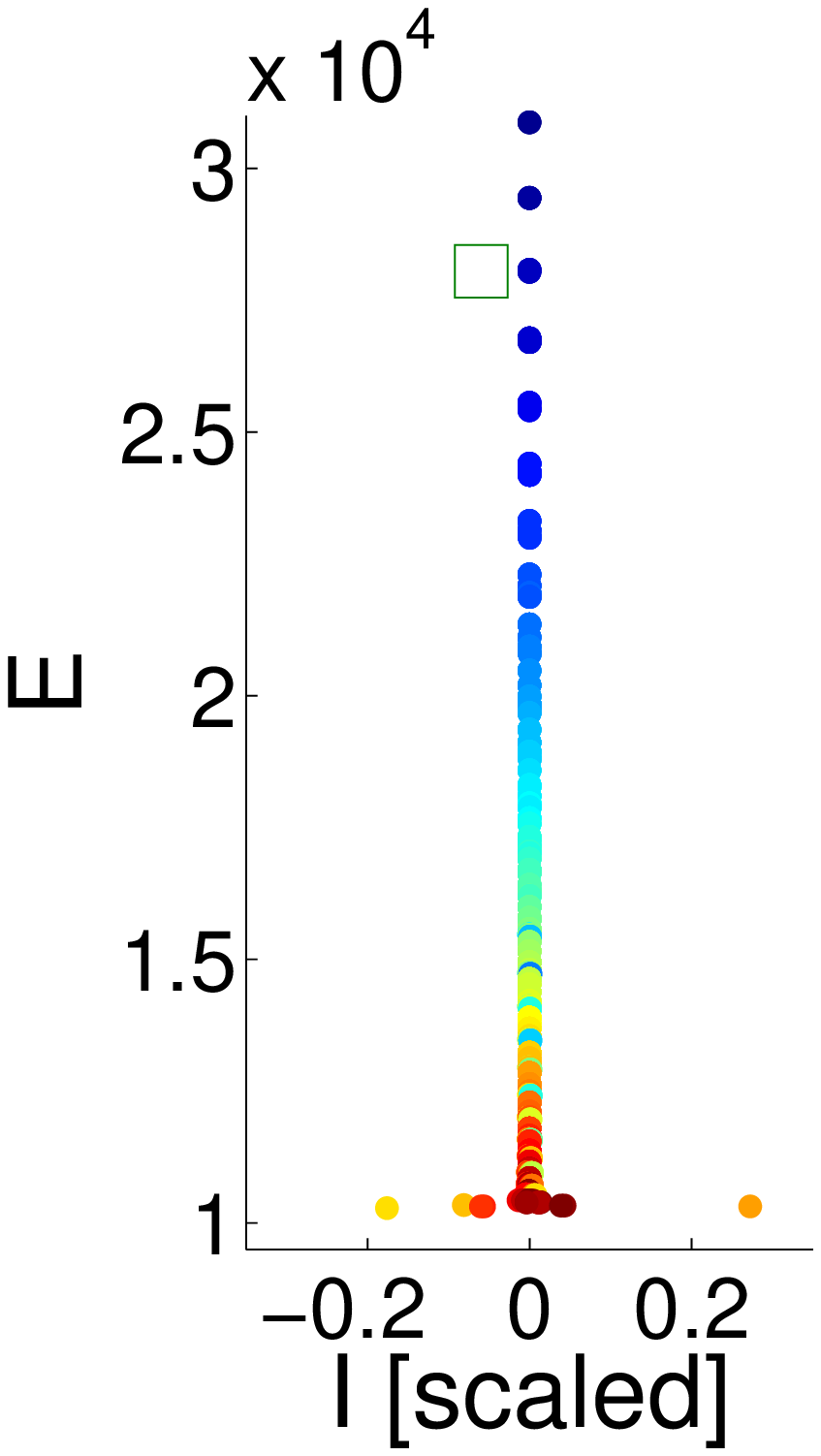}

\caption{(Color online)
The quantum spectrum of $N=42$ bosons in a triangular BEC trimer. 
Panels (a-e) are for ${(\Phi,u)}$ as follows: 
${(0.2\pi,0.2)}$; ${(0.2\pi,2.5)}$; ${(0.7\pi,2.5)}$; ${(0.6\pi,8.5)}$; ${(0.6\pi,1470)}$.
The units of time are such that $K{=}1$.
Each point represents an eigenstate color-coded by 
its purity (blue $(1/S)\sim1$ to red $(1/S)\sim3$), 
and positioned according to its energy $E_{\alpha}$ 
and its scaled current $I_{\alpha}/(NK/M)$.
\rmrk{The expected location of vortex-states, 
if they survive, is given by \Eq{e129} and \Eq{e127}, 
and represented as empty blue circles.}
In panels (b-d), due to bifurcations, 
stable solitons are found in the expected semi-classical 
locations (indicated by empty green squares). 
In (c) as opposed to (b) the rotation frequency 
is large enough to stabilize an intermediate vortex-state. 
In (e) as opposed to (d) the interaction is strong, 
and the Mott transition of the ground-state 
is reflected in its low purity.
The remaining low-purity fragmented states dwell in the 
chaotic sea and do not carry a quantized current. 
For large~$M$ we expect their dispersion in~$I$ 
to shrink further down compared with the quantized values. 
The thick black line in (a) is the microcanonical value of~$I$.    
} 

\label{f2}
\end{figure*}

\section{A rotating device}

In a {typical experimental scenario} the potential is translated along 
the circuit, and can be written as ${V(\theta-\Omega t)}$.  
For theoretical analysis it is more convenient to transform the Hamiltonian to 
a rotating frame wherein the potential is time-independent. 
The rotation is thus formally equivalent to the introduction of a magnetic flux $\Phi$ 
through the ring. To relate the flux $\Phi$ to the rotation velocity $\Omega$ it is sufficient, 
without loss of generality, to write the one-particle Hamiltonian as follows:
\beq
\mathcal{H} 
\ \ &=& \ \ \frac{1}{2\mass R^2} \left(p_{\theta}-\frac{\Phi}{2\pi}\right)^2+V(\theta)
\\
\ \ &\equiv& \ \ \mathcal{H}_0 - \Omega p_{\theta} + \const~,
\eeq
where $R$ is the radius of the ring, 
$\theta$ is the angular position coordinate, 
and $p_{\theta}$ is the conjugate angular momentum. 
In the second equality $\mathcal{H}_0$ is defined 
as the Hamiltonian in the absence of magnetic flux.
Consequently one observes that $\mathcal{H}$   
is formally the same Hamiltonian as that
of a rotating system with the implied identification 
\beq
\Omega \ \ = \ \ \left[\frac{1}{\mass R^2}\right] \frac{\Phi}{2\pi}~.
\eeq
In references \cite{Udea,Carr1,Carr2} units are set such that the pre-factor in the square brackets  is unity.  
Hence, throughout this paper ${\Phi=2\pi}$ corresponds to ${\Omega=1}$ in these references.

{We consider $V(\theta)$ consisting of $M$ deep wells, for which a tight-binding model is appropriate. }
The hopping frequency $K=1/(\mass_{\text{eff}}a^2)$
is conventionally expressed in terms of 
the lattice constant $a=2\pi R/M$ and an effective mass $\mass_{\text{eff}}$.
The magnetic flux $\Phi$ implies the vector potential $A=\Phi/(2\pi R)$. 
The phase acquired as particles hop between wells is $A \times a = \Phi/M$.
For practical purpose the relation between $\Omega$ and $\Phi$ can be re-written as 
\beq
\Omega \ \ = \ \ 
\frac{2\pi}{M^2}
\left(\frac{\mass_{\tbox{eff}}}{\mass}\right) K \Phi~.
\eeq 
Note that both $K$ and $\Omega$ have dimensions of frequency.

\section{The trimer Hamiltonian}

Few-mode Bose-Hubbard systems are experimentally accessible, highly tunable, 
and theoretically tractable by a wide range of techniques. 
Since boson number is conserved, their Hilbert spaces are of finite dimension, 
and yet their classical dynamics is non-integrable. The BHH in a rotating frame is
\be{1}
\mathcal{H} = \sum_{j=1}^{M} \left[
\frac{U}{2} a_{j}^{\dag} a_{j}^{\dag} a_{j} a_{j} 
- \frac{K}{2} \left(\eexp{i(\Phi/M)} a_{j{+}1}^{\dag} a_{j} + \text{h.c.} \right)
\right]~.
\eeq
Here $j \, \text{mod}(M)$ labels the sites of the ring, $a_{i}$ and $a_{i}^{\dagger}$ are canonical destruction and creation
operators in second quantization, $K$ is the hopping frequency, and $U$ is the on-site interaction.

As described in the previous section, the phase $\Phi$ 
reflects the rotation frequency $\Omega$ of the ring.  
Without loss of generality we assume ${\Phi\in[0,\pi]}$, and ${K>0}$, and ${U>0}$. 
Negative $K$ is the same as positive $K$ with ${\Phi\mapsto\Phi+\pi}$.
Negative $U$ is the same as positive $U$ with a flipped energy landscape (${\mathcal{H} \mapsto -\mathcal{H}}$).
Negative $\Phi$ is related to positive $\Phi$ by time reversal.

The Hamiltonian $\mathcal{H}$ commutes with the total particle number $\mathcal{N}=\sum_{i}a^{\dagger}_{i}a_{i}$, 
hence the {\em operator} $\mathcal{N}$ is a constant of motion, 
and without loss of generality can be replaced by a definite {\em number} $N$.

In a semi-classical context a bosonic site can be regarded 
as an harmonic oscillator, and one substitutes $a_j=\sqrt{\bm{n}_j}\eexp{i\varphi_j}$. 
Dropping a constant we get
\be{2}
\mathcal{H} =  \sum_{j=1}^{M} \left[
\frac{U}{2} \bm{n}_{j}^2 
- K \sqrt{\bm{n}_{j{+}1} \bm{n}_{j}} \cos\left((\varphi_{j{+}1}{-}\varphi_{j}){-}\frac{\Phi}{M}\right)
\right]~.   
\eeq 
Since $N$ is a constant of motion \Eq{e2} describes $M{-}1$ coupled 
degrees of freedoms. Accordingly the trimer (${M=3}$) is equivalent to two coupled 
pendula, featuring mixed-phase space with chaotic dynamics. 
{In practice it is convenient to define phase-space configuration 
coordinates ${r=(r_1,r_2)}$ and associate variables ${q=(q_1,q_2)}$ as follows:}
\beq
r_1=-\frac{1}{2N}(\bm{n}_3-\bm{n}_2)~, & \ \ & r_2=-\frac{1}{2N}(\bm{n}_1-\bm{n}_3)~, 
\\
q_1=\varphi_3-\varphi_2~, \ \ \ \ \ \ \  && q_2=\varphi_1-\varphi_3~.
\eeq
Note that in a {\em semi-classical} perspective $r$ and $q$ 
are canonically conjugate variables 
with commutation relation ${[r,q]=i\hbar}$, where $\hbar=1/N$.
%
%

Given the model parameter ${(\Phi, K,U,N)}$ we use standard 
re-scaling procedure (of $\bm{n}$ as described above, and of {the} time) 
in order to deduce that the classical {equations} of motion are controlled 
by two dimensionless parameters $(\Phi,u)$, 
where the dimensionless interaction strength is
\beq
u \ \ = \ \ \frac{NU}{K}~.
\eeq
Upon quantization we have the third dimensionless 
parameter~${\hbar=1/N}$.

\section{The current}

The eigenstates $|E_{\alpha}\rangle$ of \Eq{e1} 
are {characterized by the current ${I=\langle  \mathcal{I} \rangle}$ that they carry.}
The outcome of the standard definition ${\mathcal{I} \equiv -(\partial\mathcal{H}/\partial\Phi)}$ 
is gauge dependent. For the translation-symmetric gauge of \Eq{e1}  
we get the bond-averaged current.

For a {{\em single}} particle in a ring 
both $\mathcal{H}$ and $\mathcal{I}$ commute 
with the {\em non-degenerate} displacement operator $\mathcal{D}$, 
{whose eigenstates are the momentum-orbitals 
with eigenvalues $\eexp{i (2\pi/M) m}$, 
where $m$ is an integer modulo~$M$.}
Hence $\mathcal{H}$ and $\mathcal{I}$ commute with each other.
{This commutation holds also if we have $N$ 
{\em non-interacting} particles. If all the $N$ particles 
occupy the same orbital we get a ``vortex" eigenstate 
whose energy is} 
\be{126}
E_m[U{=}0] \ \ = \ \  -NK \, \cos\left(\frac{1}{M}(2\pi m-\Phi)\right)~,
\eeq
that is characterized by a definite value of current:
\be{127}
I_m \ = \ \frac{N}{M}K \, \sin\left(\frac{1}{M}(2\pi m-\Phi)\right)~.
\eeq
{This current is ``quantized", meaning that 
the scaled current $I_m/(NK/M)$ has a set of $M$ allowed values.}    
 
The Hamiltonian $\mathcal{H}$ and the current 
operator $\mathcal{I}$ no longer commute if we add 
the interactions between the particles.  
Due to the interactions {\em the current is not a constant of motion}:
the displacement operator still commutes with $\mathcal{H}$,
but decomposes it merely into $M$ blocks. 
Unlike a continuous ring system, the current~$\mathcal{I}$ cannot be identified 
with the total angular momentum. 
Still we can characterize each eigenstate of the Hamiltonian 
by its {\em average} current: 
\beq
I_{\alpha} \ = \ \langle  \mathcal{I} \rangle_{\alpha}, 
\ \ \ \ \ \ \ \mbox{$\alpha$ = eigenstate}~.
\eeq
{We note that in the presence of interaction 
the vortex-states are no longer eigenstates: They 
can at best, only {\em approximate} eigenstates. Similarly, 
a general eigenstate is not expected to be characterized 
by a quantized $I_{\alpha}/(NK/M)$. 
Instead, as explained below, we expect to obtain a relatively small scaled current rather 
than to witness ``superfluidity". } 

In \Fig{f2} we plot the numerically calculated 
spectrum of the BHH \Eq{e1} {for representative values 
of $u$  and~$\Phi$}. The points ${(E_{\alpha},I_{\alpha})}$
correspond to the eigenstates of the Hamiltonian.
{Their color indicates their one-particle coherence, 
which we define later in Section V.
The quantized current values of \Eq{e127} are marked by circles. }

In a classical context the average current 
of a microcanonical ergodic state is calculated using 
the standard statistical-mechanics prescription:
\beq
I_{cl}(E) \ = \ \frac{1}{\mathsf{A}}\int \mathcal{I}(r,q) \ \delta\left(\mathcal{H}(r,q)-E\right) \ d\vec{r}d\vec{q}~,
\eeq
where $\mathsf{A}=\int\delta\left(\mathcal{H}(r,q){-}E\right)d\vec{r}d\vec{q}$ \, is the area of the energy surface. 
%
{In the quantum context the same result (disregarding small fluctuations) 
is obtained if we average the $I_{\alpha}$ over a small energy window.
The width of the energy window should be classically small 
but quantum mechanically large: it should contain 
many eigenstates within a small energy interval. }

{The classical function $I_{cl}(E)$ has a smooth variation with respect to $E$, 
as implied by its definition. For a fully chaotic system
the semi-classical expectation would be to have microcanonical-like 
quantum eigenstates, spread throughout the energy surface such that ${I_{\alpha}\approx I_{cl}(E_{\alpha})}$, 
with very small fluctuations.  For illustration purpose this hypothesis 
is depicted by the black curve in \Fig{f2}a. }
Contrary to this expectation one observes that there is very 
large dispersion of $I_{\alpha}$ values around 
the microcanonical $I_{cl}(E)$ value. The deviation from the classical-ergodic prediction originates from either quantum interference effects or from incomplete ergodicity due to the mixed phase-space. Both issues are addressed in the following sections.

\begin{figure*}

(a) \hspace*{3.2cm} (b) \hspace*{3.2cm} (c) \hspace*{3.2cm} (d) 

\includegraphics[width=4cm]{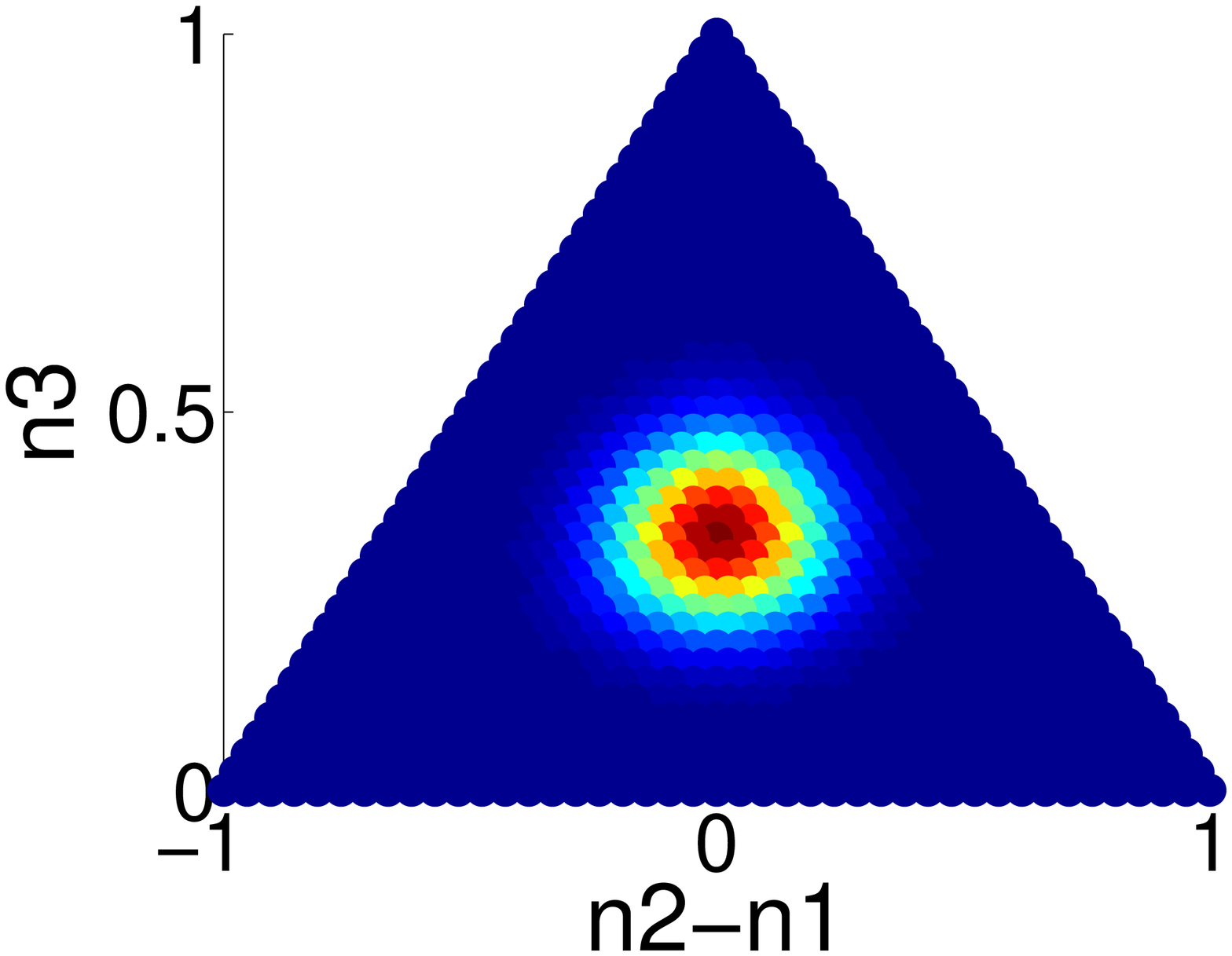} 
\includegraphics[width=4cm]{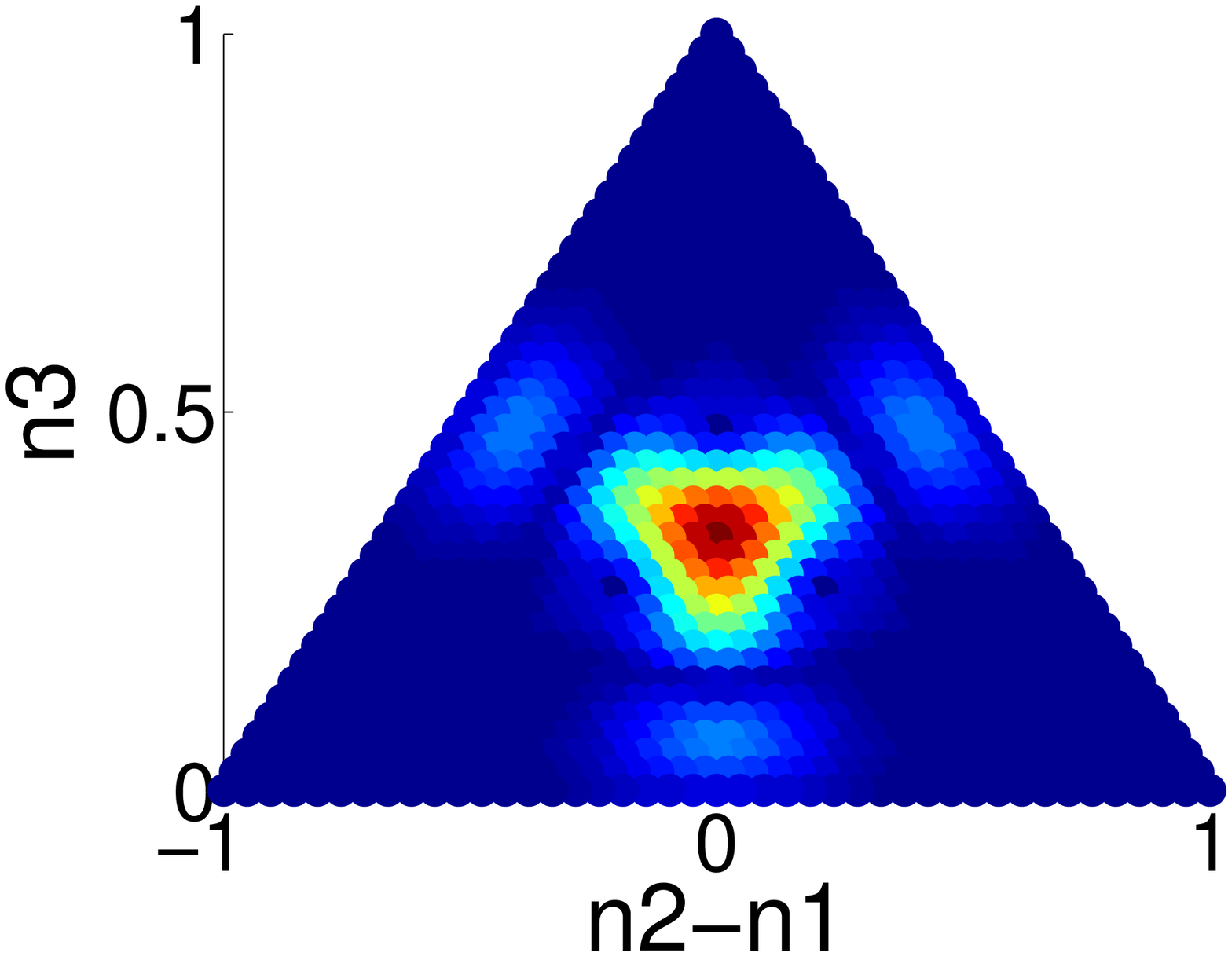} 
\includegraphics[width=4cm]{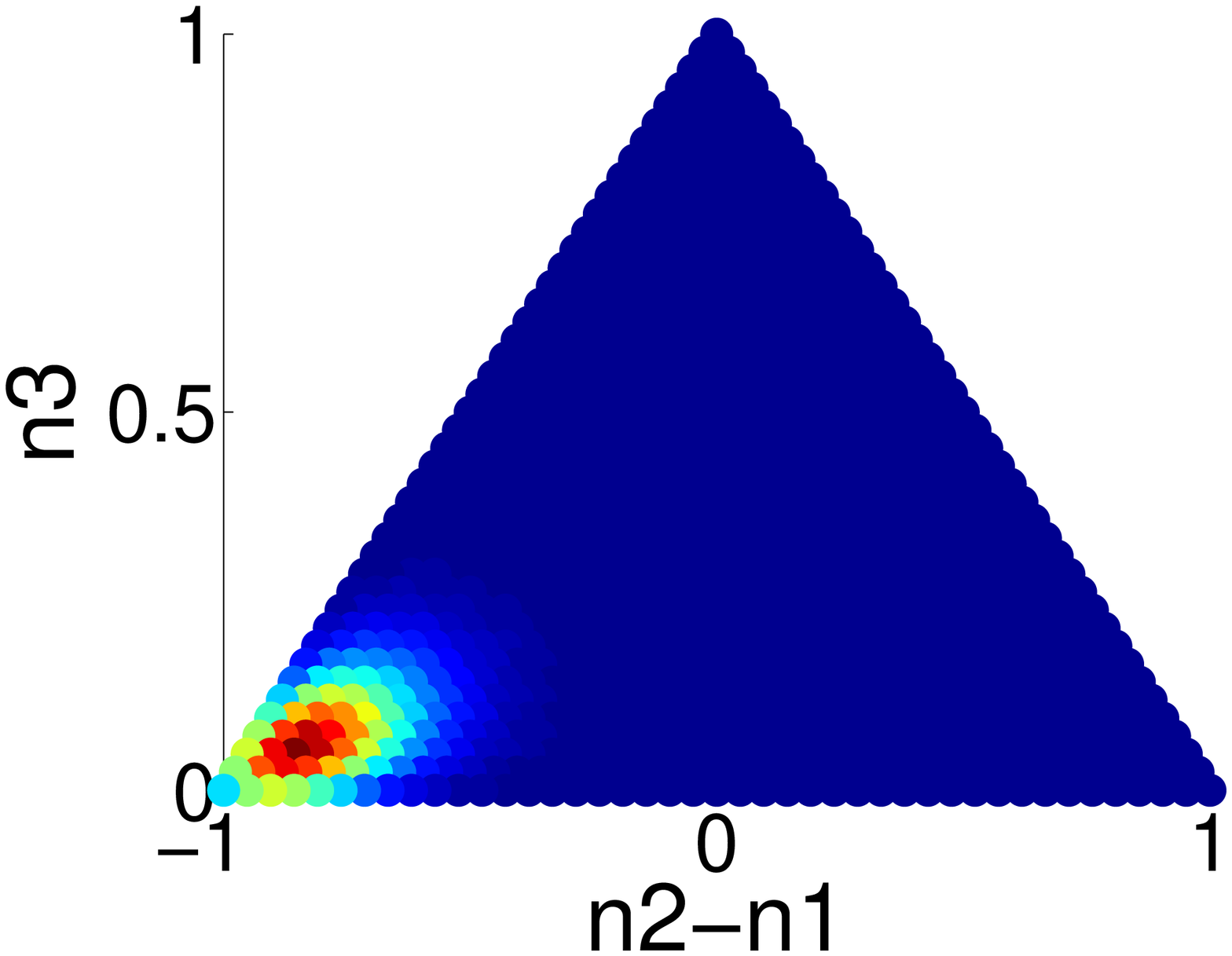} 
\includegraphics[width=4cm]{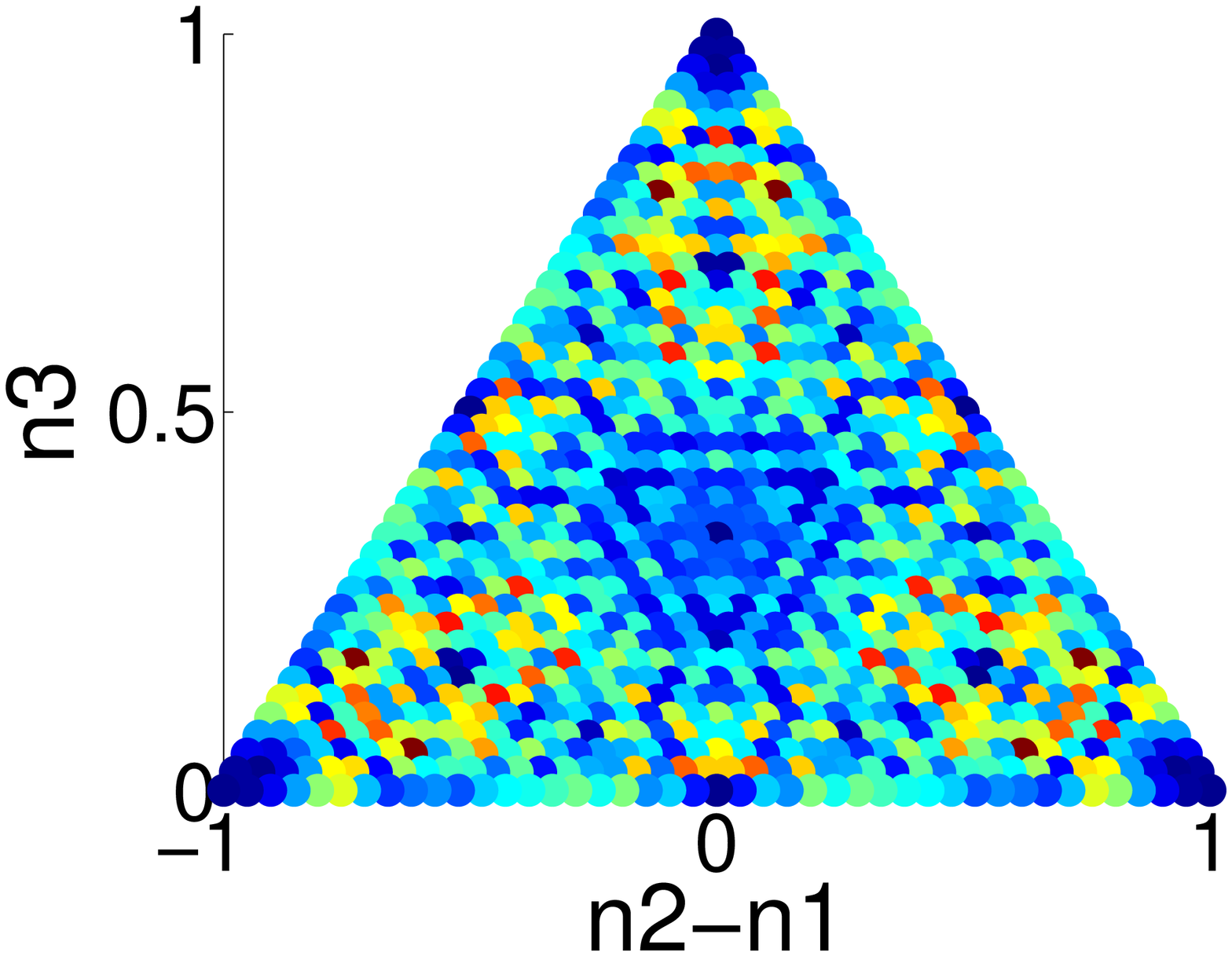} 

\caption{(Color online)
Representative eigenstates for the spectrum that is plotted in panel (c) of \Fig{f2}. 
Each panel is an image of the probability density in $r$ space, where blue and red correspond 
to zero and maximum density respectively.
(a) The ground-state vortex ${E_{\alpha} \approx -14}$.
(b) A metastable vortex ${E_{\alpha} \approx 9}$.
(c) A self trapped bright soliton ${E_{\alpha} \approx 62}$.  
(d) A low purity state in the chaotic sea ${E_{\alpha}\approx 29}$. 
{The color code is blue (low density) to red (high density).}}

\label{f8}
\end{figure*}

A few representative examples of eigenstates are shown in \Fig{f8}.  
For each eigenstate $|E_{\alpha}\rangle$ we plot the probability density in $r$ space:
\beq
|\Psi(r)|^2 \ \ = \ \ \Big|\langle r | E_{\alpha} \rangle\Big|^2~. 
\eeq
The image axes are $(\bm{n}_1{-}\bm{n}_2)/N$ and $\bm{n}_3/N$. 
{Panel~(a) displays a ground state wavefunction: 
all the particles occupy the lowest momentum-orbital, 
implying roughly equal occupation of the 3~sites.
In panel (b) the particles occupy an intermediate momentum-orbital,
while in panel~(c) they occupy mainly a single site.
Panel~(d) is an example for a roughly ergodic eigenstate, 
and should be contrasted with the non-ergodic eigenstate of panel~(b).  
The classification of the eigenstates will be further discussed
in the following sections. }

\section{Condensation and purity}

{In this section we identify  {\em condensation} with one-particle coherence, 
and clarify that a condensate corresponds to} a {\em coherent-state}, 
supported by a {\em fixed-point} of the underlying classical Hamiltonian. 
Consequently we distinguish between two types of eigenstates that resemble 
coherent-states: {\em vortex-states} and {\em solitons}.   
  
A non-fragmented condensate is formed by macroscopic occupation 
of a single one-particle orbital~$k$. 
Namely it can be written as ${(b_k^{\dag})^N|\small{vacuum}\rangle}$,   
where $b_k^{\dag}=\sum_j\alpha_j^k a_j$ creates a particle 
in some superposition of the site modes, 
with $c$-number coefficients $\alpha_j^k$.
{Such states are many-body coherent 
states in the generalized Perelomov sense \cite{CS}. 
Their phase-space representations are minimal wave-packets
situated at some point ${z=(r,q)}$  of phase-space.}

We quantitatively  characterize the fragmentation of an eigenstate by 
its single-particle purity, 
\beq
S \ \ \equiv \ \ \trc(\rho^2)~,
\eeq
where \rmrk{$\rho_{ij}=(1/N)\langle a_j^{\dag} a_i\rangle$}
is the one-body reduced probability matrix. 
Roughly speaking $1/S$ corresponds to the number of orbitals  
occupied by the bosons. The value ${S=1}$ indicates 
a coherent-state, while a low value indicates that 
the condensate is fragmented into several orbitals.

We would like to clarify why some eigenstates 
of the Hamiltonian resemble coherent states.
For this purpose recall that the quantum eigenstates 
of the Hamiltonian are semi-classically 
supported by the energy surfaces ${\mathcal{H}(r,q)=E}$.
If the energy surface is fully connected and chaotic 
one expects the ${E_{\alpha}\sim E}$  eigenstates 
to be ergodic, microcanonical-like.
A stable fixed-point of the Hamiltonian can be 
regarded as a zero volume energy surface. 
In its vicinity the dynamics looks like that of 
an harmonic oscillator. Accordingly a Planck-cell 
volume at that region can support a coherent-state.

We identify two types of nearly-coherent eigenstates:
\begin{itemize}
\item{{\em Vortex-states} are eigenstates that resemble 
a condensate in one of the single-particle momentum-orbitals 
of the ring.  
They are supported by fixed-points that are 
aligned along the ${r=0}$ axis of phase-space, 
with ${q_1=q_2=(2\pi/M)m}$. 
Vortex-states as well as their one-particle 
excitations have high purity ${S \sim 1}$.}
\item{
{\em Self-trapped states}, also known as bright solitons,
are eigenstates that resemble a condensate 
in a localized orbital.
They are supported by fixed-points that are 
generated via a bifurcation once a vortex-state
looses its stability. This bifurcation 
scenario will be analysed in Section~IX.}
\end{itemize}

The triangular trimer Hamiltonian always has 
at least two {stable} fixed-points:
one that corresponds to the lowest energy, 
and one that corresponds to the upper-most energy.  
Accordingly both the ground-state and the upper-state 
are coherent  in the large $N$ limit. 
{Note that the upper-state can be regarded 
as the ground-state of the BHH with attractive interaction (${U\mapsto -U}$) 
as discussed in the paragraph that follows \Eq{e1}.}

The intermediate energy surfaces have a large area, 
hence microcanonical-like states located 
there are not coherent. 
{However, with a mixed-phase space more fixed-points
can be found at intermediate energies. A major finding 
of this work is that hyperbolic (unstable)
fixed-points can support {\em meta-stable} coherent states.}

\begin{figure*}

(a) \hspace*{3.2cm} (b) \hspace*{3.2cm} (c) \hspace*{3.2cm} (d) 

\includegraphics[width=4cm]{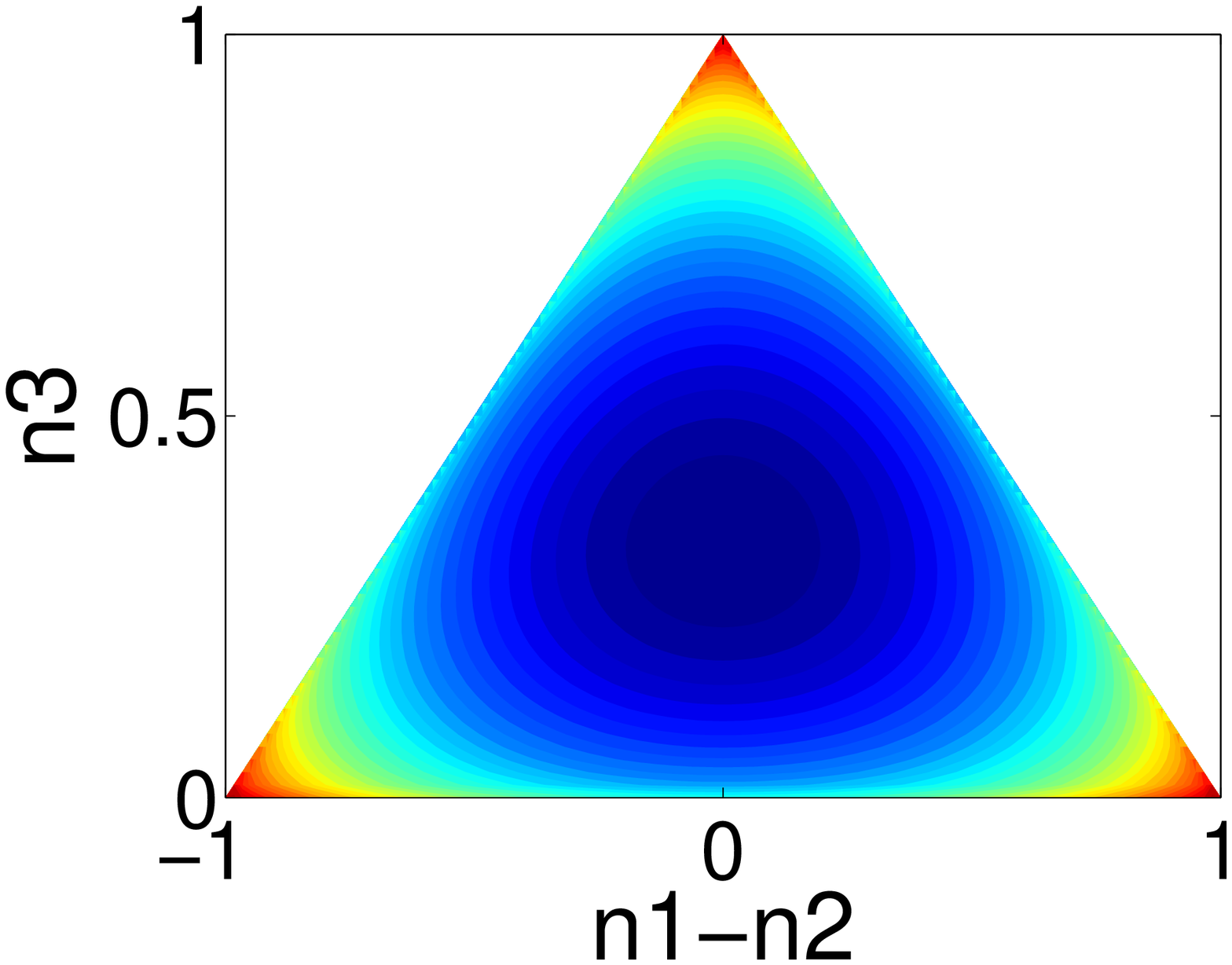}
\includegraphics[width=4cm]{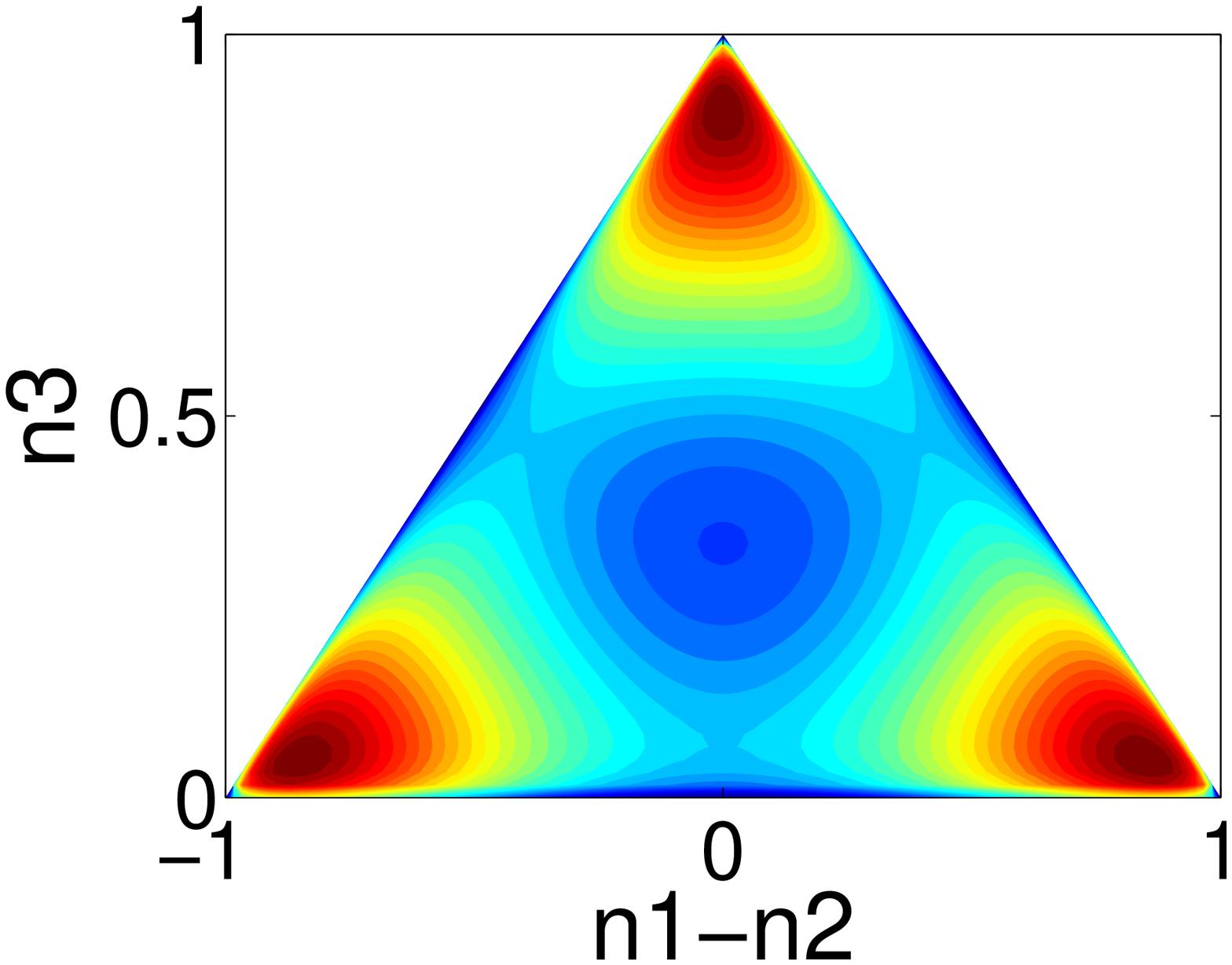}
\includegraphics[width=4cm]{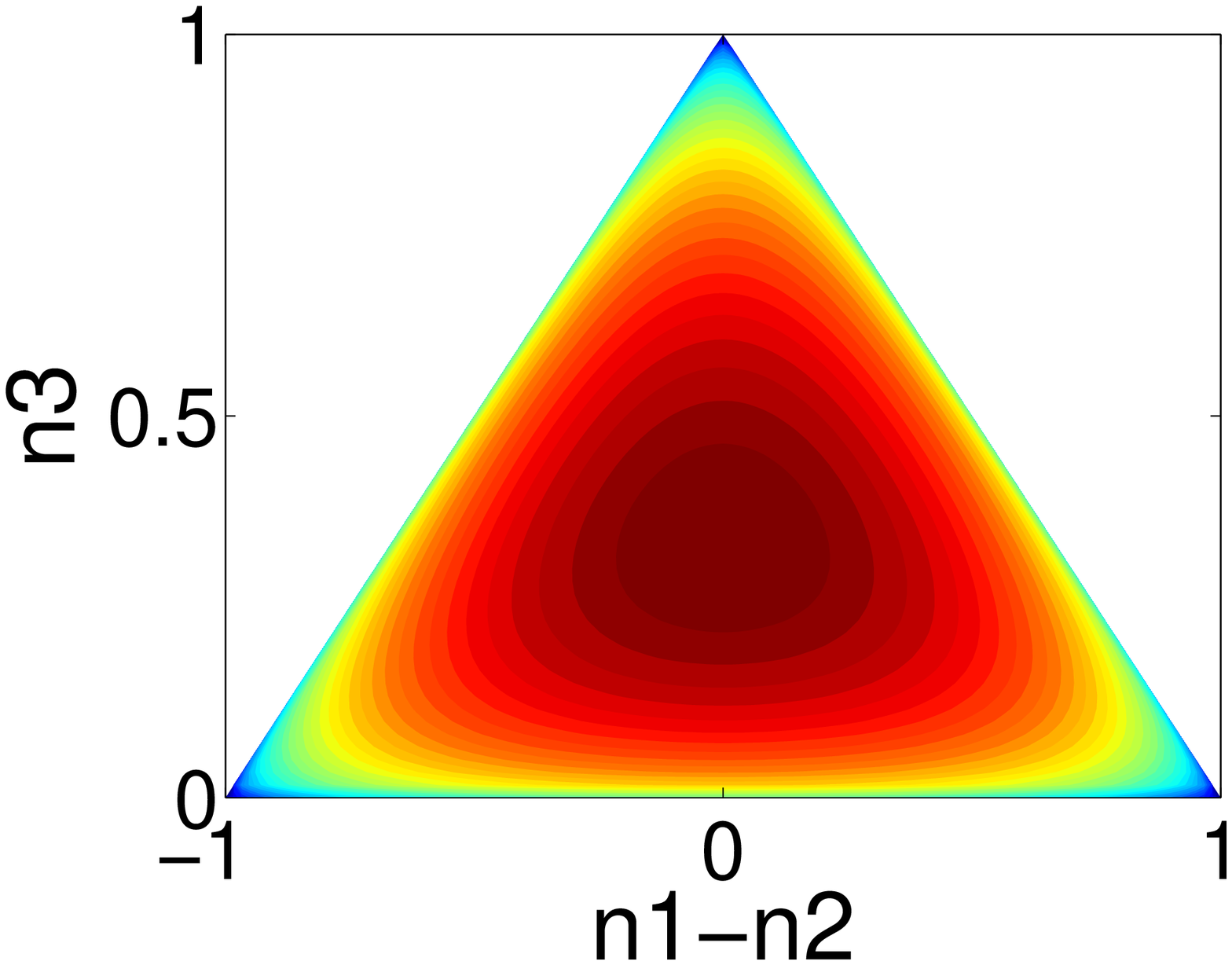}
\includegraphics[width=4cm]{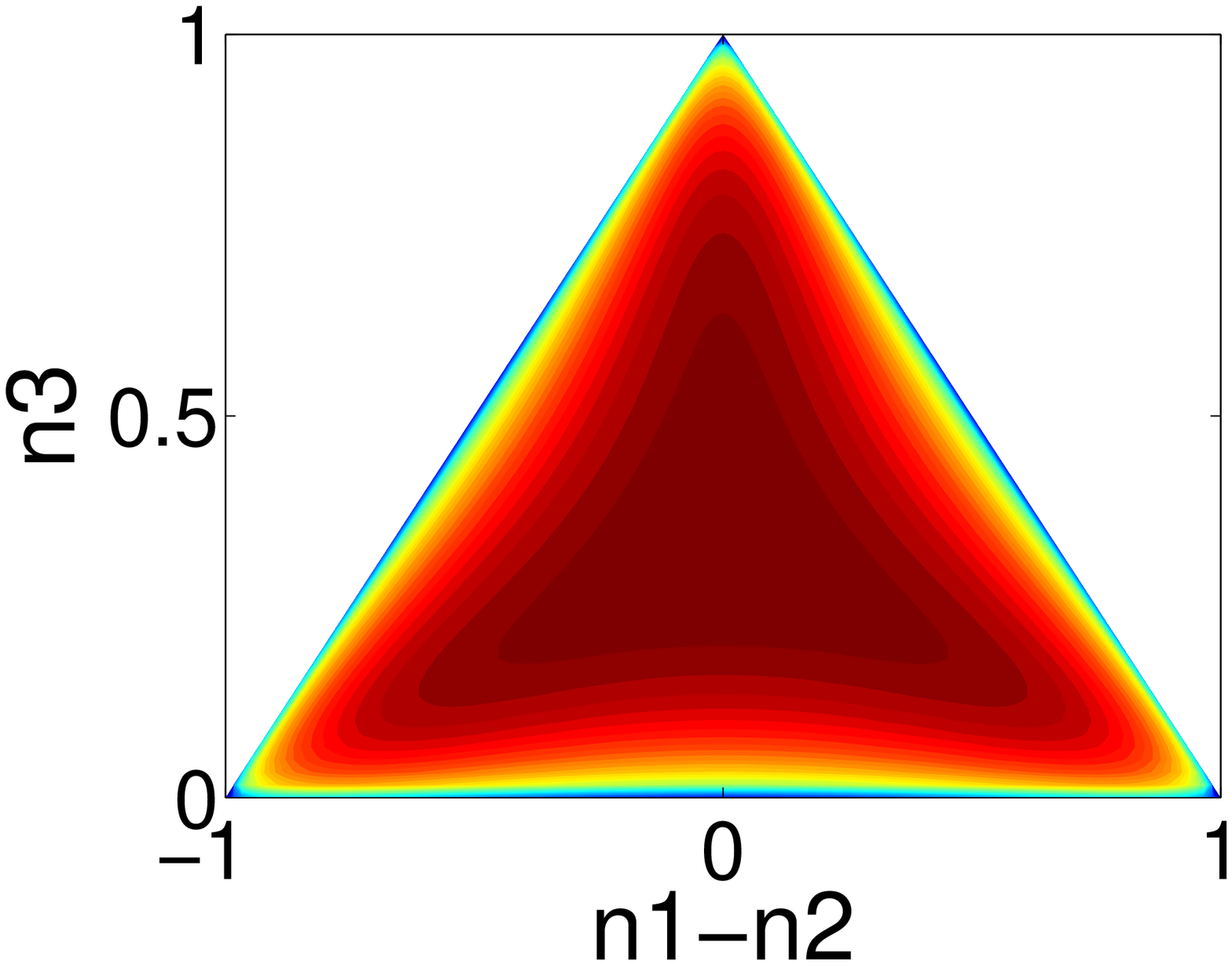}

\caption{(Color online)
Images of the energy surfaces $V_m(r)$. 
Panels~(a) and~(b) are the bottom \rmrk{($V_0$)} and the top \rmrk{($V_{+}$)}
energy surfaces for ${(\Phi,u)= (\pi/4,2)}$, 
illustrating regime (d) in \Fig{f4}. 
Panels~(c) and~(d) are the top energy surfaces 
for ${(0.8\pi,0.4)}$ and ${(0.8\pi,1.8)}$, 
illustrating regimes (a) and (b) in \Fig{f4}. 
The color code is blue (low energy) to red (high energy).
See \Fig{f3} for a section along the symmetry line. 
}

\label{f6}
\end{figure*}

\section{The energy spectrum}

The $E_{\alpha}$ spectrum is plotted in \Fig{f2} for several representative parameter sets.
Eigenstates are classified by their {\em current}~($I$), 
and color-coded according to their {\em one-particle purity}~($S$).   
In \Fig{f8} we display representative examples of eigenstates:      
the ground-state vortex; a metastable vortex; 
a self trapped bright soliton; and a low purity microcanonical-like state in the chaotic sea. 
Inspecting \Fig{f2} and \Fig{f8}, 
and comparing with the standard picture of \Fig{f1},  
we observe the following:  

(A) {\em Mott-Insulator transition}: In panels~(a)-(d) of \Fig{f2} the ground-state 
is a vortex-state that carries the expected quantized 
current of \Eq{e127}. The expected quantized values 
are indicated in the figures by blue circles. 
The corresponding  ground state wavefuntion is imaged in \Fig{f8}a.
The ground state retains its purity up to an extremely 
large value of~$u$. 
Beyond a critical value of~$u$
the purity of the ground-state is lost. 
This is evident from the color change of the ground-state from blue in 
panels 2(a)-2(d) to red in panel~2(e).
  
(B) {\em Self-trapping}: For small $u$, there is a stable  
vortex-state at the top of the energy landscape, 
with the expected quantized current of \Eq{e127}, 
See e.g. the highest energy current-carrying eigenstate in panel~2(a). 
If $u$ is large enough, the upper vortex-state
bifurcates, and is replaced by 3 self-trapped solitons 
carrying very little current. See e.g. the self trapped 
states in panel~2(b), that are represented 
by 3 overlapping points at the top of the energy landscape, 
\rmrk{close to the classically expected position marked by the green box.}
The wavefunction of a self-trapped state is imaged in \Fig{f8}c.

(C) {\em Metastable vortex states:} 
Looking at panel~(c) of \Fig{f2} we see 
that it is feasible to have an additional 
high purity vortex-state in an intermediate energy range
\rmrk{(here around $E\sim10$)}.
Contrary to the naive expectation, this state has 
not been mixed with the surrounding continuum, 
and it caries the expected quantized current of \Eq{e127}.
A representative wavefunction of such state is imaged in \Fig{f8}b.

(D) {\em Chaotic eigenstates}: It should be emphasized, 
as evident from the continuum of red color points in all panels of \Fig{f2}, 
that the majority of eigenstates 
are highly fragmented and are not characterized by a well-defined quantized value of current. 
The wavefunction of one such state in imaged in \Fig{f8}d.
 
Below we provide a semi-classical interpretation 
of the above findings, and deduce a schematic diagram 
of the $(\Phi,u)$ regimes. Self trapping is discussed in Section IX,  
The Mott transition in Section X, and metastability in Section XI.

\begin{figure*}

\begin{minipage}{7cm}
\includegraphics[width=\hsize]{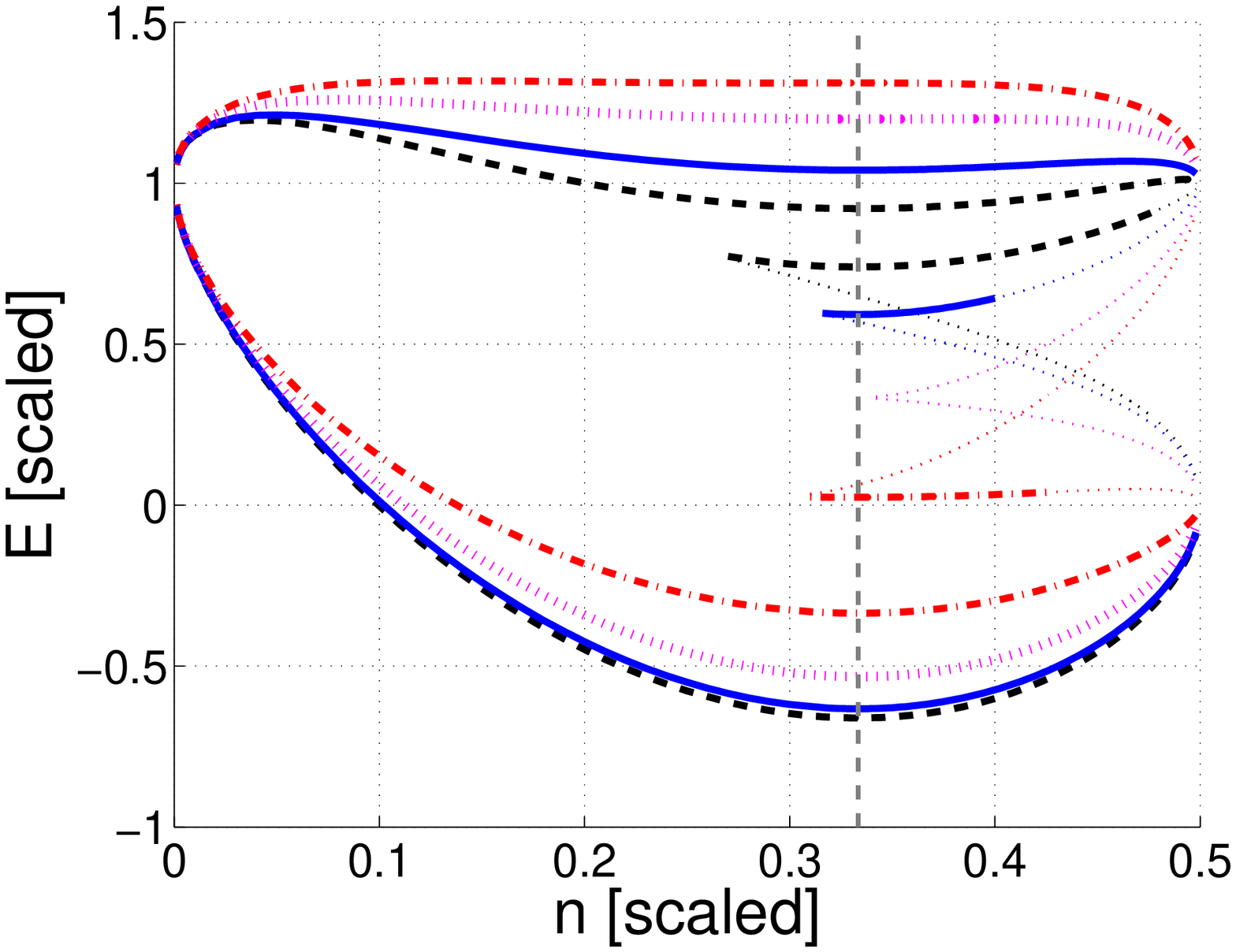}

\caption{(Color online)
A section along the symmetry line ${n_1{=}n_2{=}n}$, 
for ${u=2}$ and ${\Phi=\pi/4}$ (solid~line), 
as well as for ${\Phi=0.1\pi}$ (dashed), 
and $0.5\pi$(dotted), 
and ${0.8\pi}$ (dashdot).
Each section shows the \rmrk{top} surface and also the 
\rmrk{bottom} and intermediate surfaces, 
which are formed of extremal points. 
The thin dotted segments are formed of saddle points. 
As $\Phi$ becomes larger the intermediate surface $V_{-}(r)$ goes down in energy. 
For ${\Phi<\pi/2}$ it is formed of maxima, and its area shrinks to zero as $\Phi$ is increased.
For ${\Phi>\pi/2}$ it becomes a surface of minima, and its area expands back.
} 

\label{f3}
\end{minipage}
\hspace{10mm}
%
\begin{minipage}{9cm}
\includegraphics[width=\hsize]{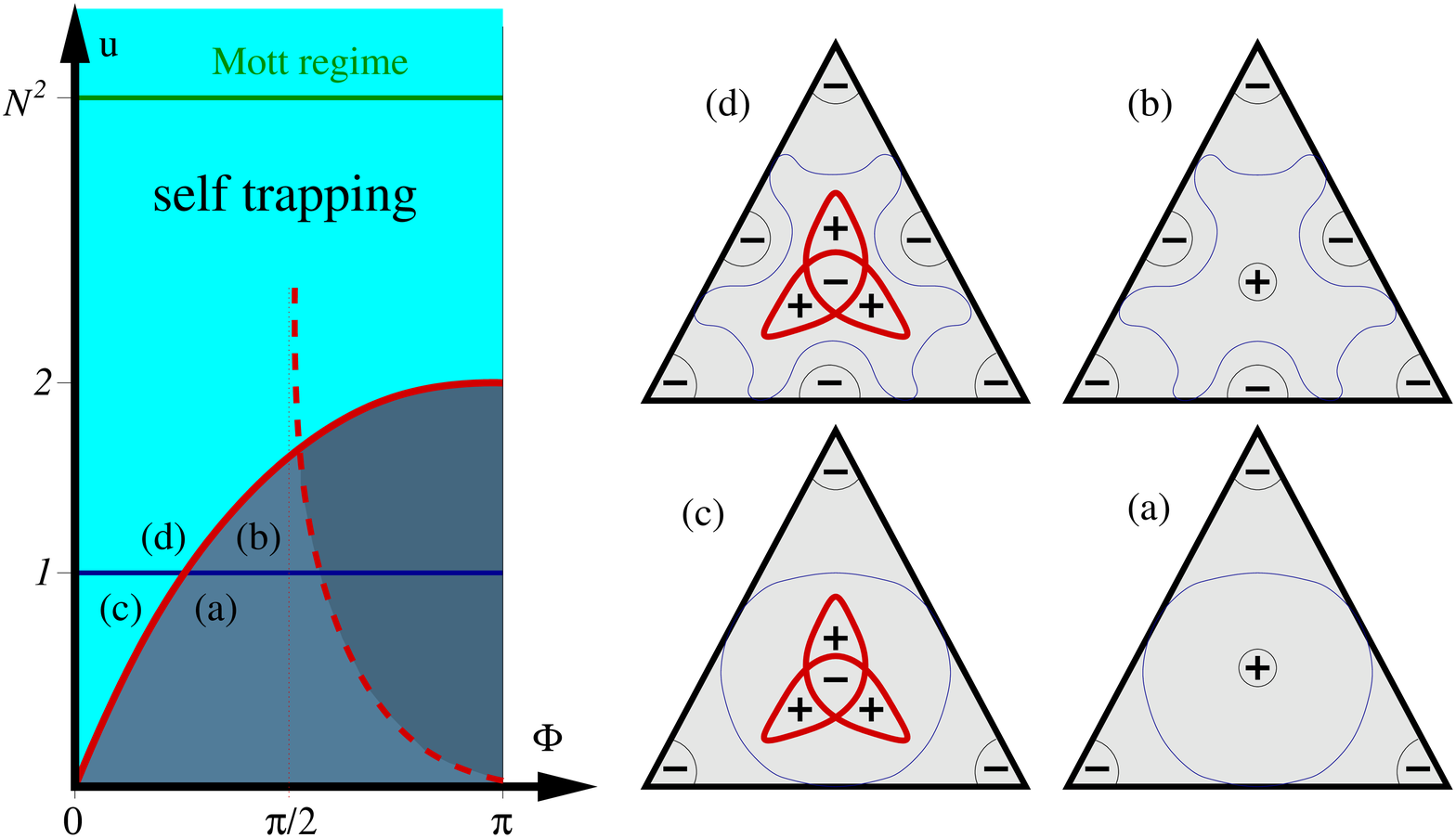}

\caption{(Color online)
Regime diagram of the triangular BEC trimer. 
The model parameters are $(\Phi,u)$.
The thick solid lines divide the diagram 
into four quarters (a-d). In each quarter 
the topography of the upper energy surface 
is different, as schematically illustrated 
on the right. The thick dashed line 
indicates the classical stability threshold 
of the intermediate energy surface. 
In the quantum analysis we observe 
quasi-stability as $\Phi=\pi/2$ is crossed: 
see the text for details and \Fig{f5} 
for demonstration.} 

\label{f4}
\end{minipage}

\end{figure*}

\section{Stability analysis}

The BHH formally describes a set of coupled oscillators. 
Schematically we can write the Hamiltonian as $H(z)$ 
with ${z=(r,q)}$ and ${\hbar=1/N}$. 
This Hamiltonian has $M{-}1$ freedoms. 
A stable fixed-point $z_0$ can support a coherent-state
provided $\hbar$ is small enough. 
  
At $r=0$ we always have 3 fixed-points 
that correspond quantum-mechanically to condensation 
in one of the 3~momentum~orbitals of the trimer.
The questions are: {\bf (i)}~whether 
these fixed-points are stable;  
{\bf (ii)}~whether they can support 
a coherent quantum state. In this section 
we discuss the first question.
In  Section~X (Mott transition) 
and in Section~XI (Metastability) 
we discuss the second question, 
which is related to having a finite~$\hbar$. 

Fixed-point stability 
is determined by linearization of the Hamiltonian 
in its vicinity, resulting in a set of 
Bogolyubov de-Gennes (BdG) equations.
This set gives $M{-}1$ frequencies $\omega_k$ 
for the Bogolyubov excitations. 
Frequencies of different signs imply {\em thermodynamic} 
instability, as occur in Hamiltonian 
of the type ${H= \omega_1 n_1 + \omega_2 n_2}$ 
with ${\omega_1>0}$ and ${\omega_2<0}$. 
Complex frequencies indicate {\em dynamical} 
instability (hyperbolic fixed-point), 
as occur in Hamiltonian ${H= p^2-x^2}$.
{We use here the same terminology as in \cite{Udea}.}

Consider a fixed-point that becomes 
thermodynamically unstable as a result of varying 
some parameters in the Hamiltonian, 
or due to added disorder.  
This means that the island that had surrounded 
the fixed-point is now opened. 
In quantum terms one may say that the 
former vortex-state can mix with the 
a finite density of zero-energy excitations, 
leading to a low purity, possibly ergodic 
set of eigenstates.


In {the} remaining part of this section we clarify how the BdG stability analysis 
is related to the Landau criterion for the stability of a superfluid motion, 
and mention the known result for an ${M\gg1}$ ring. 

The standard presentation of the Landau criterion takes the liquid 
as the frame of reference, with the walls moving at some velocity~$\Omega$. 
It is then argued that energy cannot be transferred from the walls to the liquid 
if the excitation energies satisfy ${\omega_k^{(0)} > \Omega k}$ 
for any wavenumber~$k$ of the excitation.    
In the case of phonons (${\omega_k^{(0)}=ck}$) 
this implies that $\Omega$ should be smaller 
than the speed of sound~$c$. 

It is conceptually illuminating to write the Landau criterion in the 
reference frame where the walls are at rest, and the Hamiltonian 
becomes time independent. {In this frame the superfluid is 
rotating with frequency ${\Omega_m=-\Omega}$}. The Landau conditions 
takes the form ${\omega_k>0}$ for any $k$, where 
\beq
\omega_k \ \ = \ \ \omega_k^{(0)} + \Omega_m k, 
\ \ \ \ \ \ \mbox{[standing device]},
\eeq
are the excitation energies of the vortex-state. 
For an $M\gg1$ ring, taking the continuum limit, 
the Landau criterion reasoning implies that 
the excitation energies of the $m$th vortex-state 
in a rotating device are 
\be{274}
\omega_k \ \ = \ \ \sqrt{\left(\epsilon_k+ \tilde{u} \right)\epsilon_k} \ - \ (\Omega-\Omega_m)k~,
\eeq
where $\Omega=\Phi/(2\pi)$ is the scaled rotation frequency of the device, 
and $\Omega_m=m$ is the quantized rotation frequency of the superfluid.
The integer~$k$ is the wavenumber of the Bogolyubov excitation, 
while $\epsilon_k=(1/2)k^2$ is the unperturbed single-particle energy,
and $\tilde{u}$ is the appropriately scaled interaction.   
The derivation of the first term in \Eq{e274} is standard 
and can be found for example in \cite{FV}. 
For small~$\epsilon_k$ it is common to use a linear approximation $\omega_k^{(0)}\approx ck$ 
where ${c=\sqrt{\tilde{u}/2}}$ is identified as the sound velocity.
The second term in \Eq{e274} is implied by the Galilean transformation 
that has been discussed in the previous paragraphs.
The implications of \Eq{e274} on the stability 
of the vortex-states is illustrated in \Fig{f1}.

It should be clear that the Landau criterion is not applicable 
in the case of a finite $M$ system, because one cannot 
use {a} Galilean transformation to relate $\omega_k$ to $\omega_k^{(0)}$. 
Therefore we have to utilize the phase-space picture of the dynamics in 
order to determine the regime diagram of the rotating trimer.
Whenever a vortex-state looses its stability, 
irrespective of the nature of the Bogolyubov excitations, 
we understand that superfluidity is lost.

The ``standing walls" formulation of the Landau-criterion 
makes transparent the connection to the Fermi-golden-rule picture (FGRP) 
and to the semi-classical picture (SCP).      
In the FGRP {the vortex-state is perturbed} by  walls, 
or optionally by some weak disordered potential.
{This perturbation} induces first-order coupling 
of the vortex-state to its Bogolyubov one-particle excitations.
Accordingly, in the FGRP language the Landau condition is phrased as the requirement 
of not having Bogolyubov excitations with the same energy as the vortex-state.
In the SCP, {instead of perturbing the potential}, 
one considers a weak perturbation of the vortex-state.
{This is like launching several trajectories in the vicinity of 
the fixed-point, forming an evolving phase-space distribution. The motion is stable 
if the fixed-point is a local minimum or a local maximum. 
If the fixed-point is unstable the phase-space distribution deforms
and spreads over a large region of the energy shell:
consequently the current~$I$ is diminished. 
Technically, this linear-stability analysis leads 
to the BdG equations discussed 
in the beginning of this section.
As pointed out, the eigenvalues of the BdG equations 
determine whether the vortex-state is stable or not, 
leading in the case of translationally-invariant ring to the Landau criterion.}

If the vortex-state is meta-stable, then quantum tunnelling or thermal activation 
are required in order to get a ``phase slip" to a lower vortex-state. 
This goes beyond the ``Landau criterion", but still can be addressed 
using the SCP, possibly combined with FGRP 
and optionally using WKB-type approximation.

\begin{figure*}
\includegraphics[width=4.5cm]{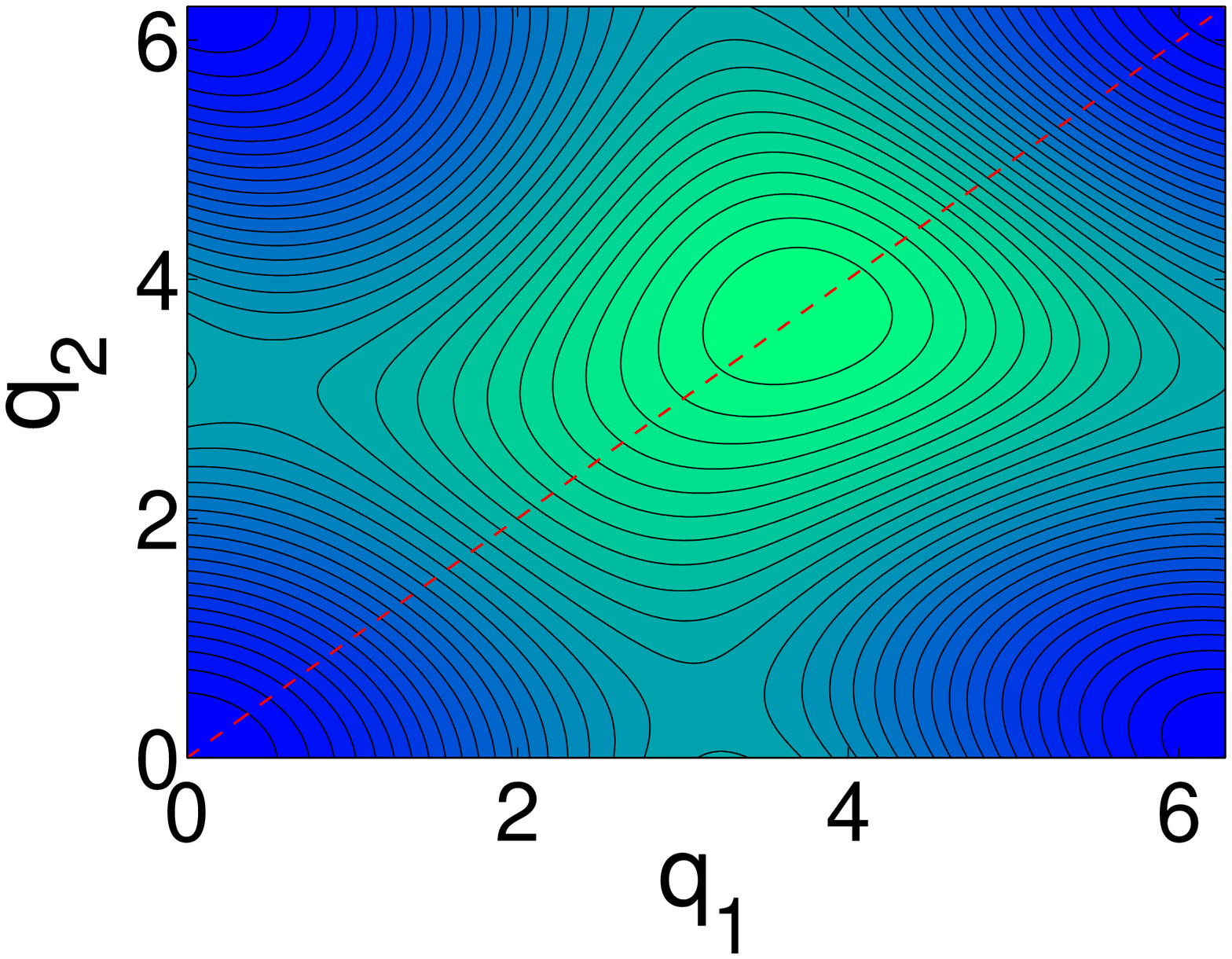}
\ \ \ \ 
\includegraphics[width=4.5cm]{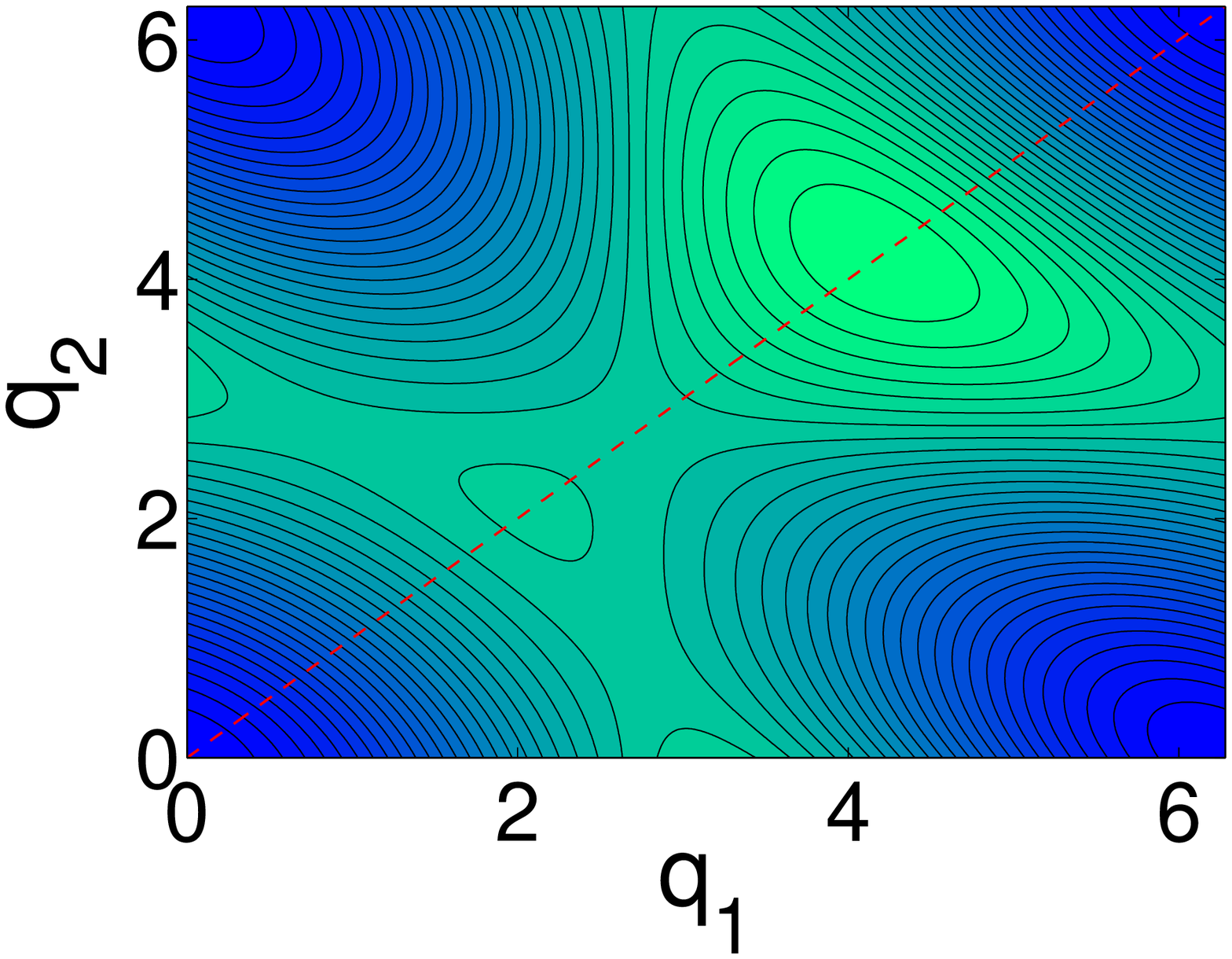}
\ \ \ \ 
\includegraphics[width=4.5cm]{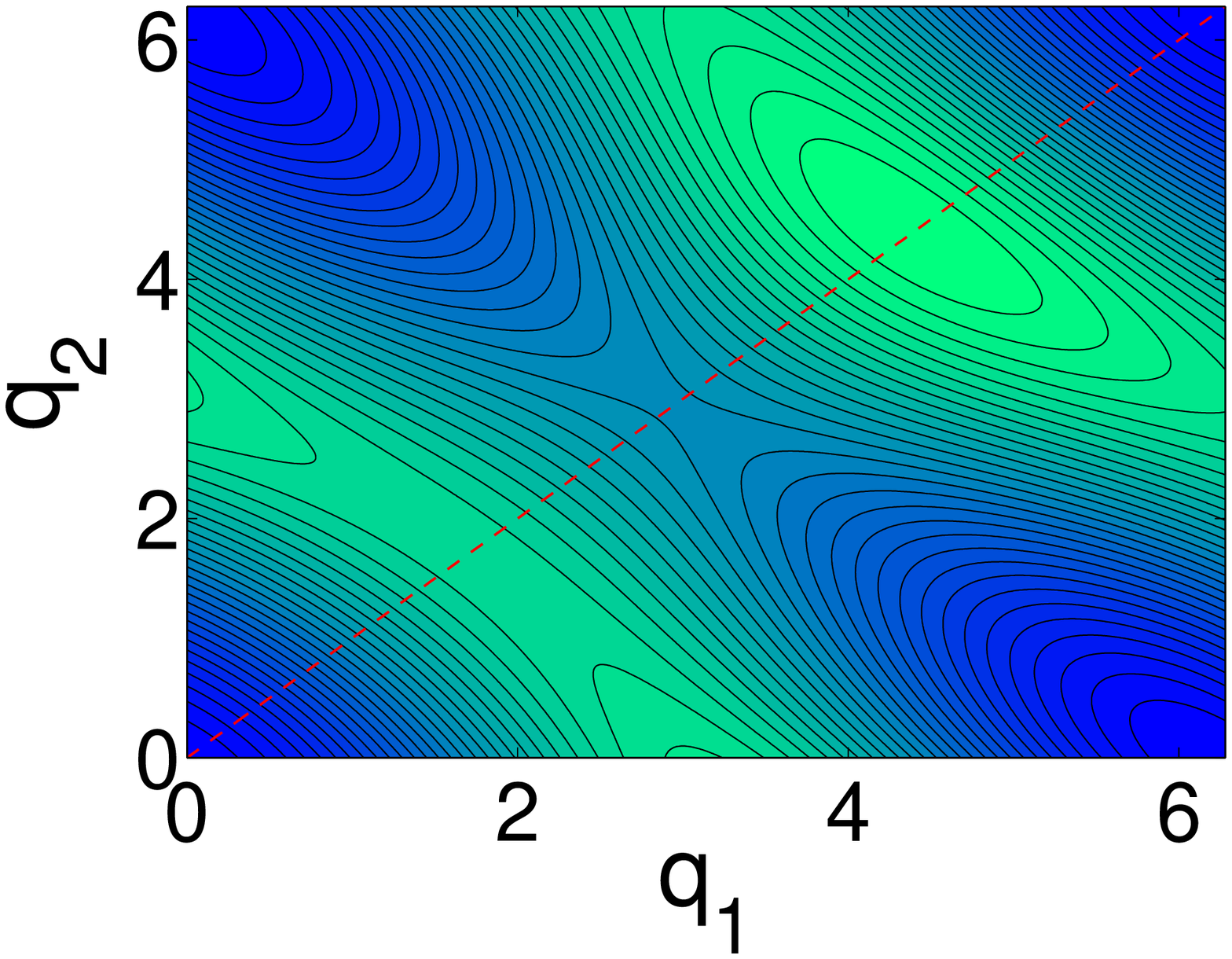}

\caption{(Color online)
Images of $E=\mathcal{H}(r,\varphi)$ as a function of $q_1$ and $q_2$ 
for ${\Phi=0.2\pi}$. In the three panels $n_1{=}n_2{=}n$, 
hence the fixed-points of interest are along the symmetry line ${q_1=q_2}$. 
The left panel is at ${n=(1/10)N}$,  
and it has a single minimum and a single maximum.   
The middle panel is at the symmetry point ${r_0=0}$, 
where ${n=(1/3)N}$,  
and it has a single minimum and two maxima.  
Consequently in the vicinity of $r_0$ we have three energy surfaces.
The right panel is at ${n=(9/20)N}$,
and it has again a single minimum and a single maximum.  
} 

\label{f7}
\end{figure*}

\section{The regimes diagram}

We take ${r}$ to be a triangular configuration 
space. It consists of points ${(\bm{n}_1,\bm{n}_2,\bm{n}_3)}$,
such that $\bm{n}_1+\bm{n}_2+\bm{n}_3=N$. 
The phase differences ${q}$ correspond to 
the conjugate momenta: they determine the velocity~$\dot{r}$.  
The energy landscape of the Hamiltonian can be visualized 
using the Peierls-Nabarro surfaces $V_m(r)$, formed of its extremal points 
under phase variation \cite{Henn1,Henn2,Kivshar93,Rumpf04}. 
Lower Peierls-Nabarro surfaces are thus defined as 
\beq
\label{lowsur}
V(r) \ \ \equiv  \ \ \min_{\varphi}[\mathcal{H}(r,\varphi)]~,
\eeq
whereas upper Peierls-Nabarro surfaces are 
defined with ${\max[\cdots]}$ instead of ${\min[\cdots]}$.
Additionally we may have pieces of surfaces 
that consists of saddle points. 
 
The Peierls-Nabarro surfaces of the 
triangular trimer are shown in \Fig{f6} (images) 
and in \Fig{f3} (sections).  
We have 3 Peierls-Nabarro surfaces:
{\bf (i)} a bottom surface $V_0(r)$ which is always a ``lower surface" (i.e. satisfies Eq.~\ref{lowsur}), 
{\bf (ii)} a {top} surface $V_{+}(r)$ which is always an ``upper surface", 
and {\bf (iii)} an {intermediate} surface $V_{-}(r)$, 
which is either an ``upper" or a ``lower" surface depending on~$\Phi$:  
for ${\Phi<\pi/2}$ it is an ``upper" surface that is formed of local maxima,  
whereas for ${\Phi>\pi/2}$ it is a ``lower" surface.  
\rmrk{The various curves in \Fig{f3} illustrate how the intermediate 
surface $V_{-}(r)$ is modified as a function of the rotation frequency: 
As $\Phi$ becomes larger this surface goes down in energy, 
and its area shrinks to zero at ${\Phi=\pi/2}$.
As $\Phi$ is increased further its area expands back.}

A stable fixed-point is either a minimum of a ``lower" surface 
or a maximum of an ``upper" surface. 
Note also that the upper-most fixed-point 
can be envisioned as the ground-state of 
the $U\mapsto -U$ Hamiltonian.   
  
The diagram of \Fig{f4} summarizes the different parametric 
regimes of the model: each regime is characterize by a different 
type of $V_m(r)$ topography. 
The central point ${r_0=0}$ is a fixed-point
of the 3 surfaces, with energies $E_{m}=V_{m}(r_0)$. 
The fixed point $E_0$ is always stable.  
In contrast, the $E_{\pm}$ fixed-points may be stable or unstable, 
depending on their curvature $V_{\pm}''(r_0)$,  
where prime denotes differentiation in the ``radial" direction.

The fixed-points are situated on the symmetry lines 
in $r$ space. So we can restrict the analysis 
along, say, ${\bm{n}_1=\bm{n}_2=n}$, hence ${\bm{n}_3=N{-}2n}$, and 
\be{327}
&&\mathcal{H}(r,q) =  
\frac{U}{2}\left(\bm{n}_1^2+\bm{n}_2^2+ \bm{n}_3^2 \right) 
\\ \nonumber
&& \ \ \ \ \ \ 
- K \Big(  \sqrt{\bm{n}_2\bm{n}_3} \,\cos\left(q_1-\scriptstyle{\frac{\Phi}{3}}\right) +  \sqrt{\bm{n}_3\bm{n}_1} \,\cos\left(q_2-\scriptstyle{\frac{\Phi}{3}}\right) 
\\ \nonumber
&& \ \ \ \ \ \ \ \ \ 
+ \sqrt{\bm{n}_1\bm{n}_2} \,\cos\left(q_1+q_2+\scriptstyle{\frac{\Phi}{3}}\right) \Big)~.
\eeq
The lowest surface ($m=0$) has a trivial topography. 
In particular for ${\Phi=0}$ the lower surface is 
\beq
V_{0}(r) &=& 
\frac{U}{2}\left(\bm{n}_1^2+\bm{n}_2^2+ \bm{n}_3^2 \right) 
\\ \nonumber
&& - K \, \left(  \sqrt{\bm{n}_2\bm{n}_3} +  \sqrt{\bm{n}_3\bm{n}_1}  + \sqrt{\bm{n}_1\bm{n}_2} \right)~.
\eeq
For general $\Phi$ the extremal values (at a given $r$ location) are still 
situated along the line $q_1=q_2$. 
{This is clear by inspection of \Eq{e327}, which is illustrated in \Fig{f7}:}  
Depending on $\bm{n}$ and $\Phi$ we have either a single minimum and a single maximum,
or two minima and a maximum, or a minimum and two maxima.
It follows that one has to find the 3 extremal points $q_m(n)$ of the function
\be{11}
&& H(q;n) \ = \  \frac{1}{6}N^2U + 3U \left(n-\frac{N}{3}\right)^2  
\\ \nonumber
&& \ \ \ \ - K \left(2\sqrt{(N-2n)n} \,\cos\left(q-\scriptstyle{\frac{\Phi}{3}}\right) + n \,\cos\left(2q+\scriptstyle{\frac{\Phi}{3}}\right) \right)~. 
\eeq
Then one obtains a section of the surface $V_m(r)$ 
along the principal ``radial" direction ${\bm{n}=(n,n,N{-}2n)}$
\beq
V_m(n) \ \ = \ \ H(q_m(n);n)~. 
\eeq
The most interesting fixed-points of $V_m(r)$ 
are situated at the central point ${\bm{n}=(N/3,N/3,N/3)}$,  
for which ${n=N/3}$. The three $r_0$ fixed-points 
support the vortex-states. 
The energies of the fixed-points are:
\be{129}
E_{m} =  V_m(r_0)  = \frac{1}{6}N^2U -NK \ \cos\left(\frac{2\pi m - \Phi}{3}\right) \ ~,
\eeq
with ${m=0,\pm 1 }$.
{Note consistency with \Eq{e126} and the associated expression for the current \Eq{e127}.} 
A lengthy but straightforward calculation leads to the following 
result for the curvature at~$r_0$. This is required in order to 
determined whether they are stable or not:
\be{6}
&& V_{\pm}''(r_0) \ \ = \ \ \left.\frac{d^2 H(q_{\pm}(n),n)}{dn^2}\right|_{N/3} 
%
%
\\ \nonumber
&& \ \ = 6\frac{K}{N} \left[u 
- \frac{6 - 9\cos\left(\frac{\pi\pm2\Phi}{3}\right)  - 3\cos\left(\frac{\pi\mp4\Phi}{3}\right) }
{6\cos\left(\frac{\pi\mp\Phi}{3}\right) - 2\cos\left(\Phi\right) } \right]
~. \ \ \ \ \ \ 
\eeq
We shall use this result in the subsequent 
discussion of self-trapping and meta-stability.

\begin{figure*}

(a) Current $\langle I \rangle_{+}$ 
\hspace{2.5cm} (b) $1/S_{+}$  
\hspace{4cm} (c) $1/S_{+}$  
\hspace{1cm}

\includegraphics[width=5cm]{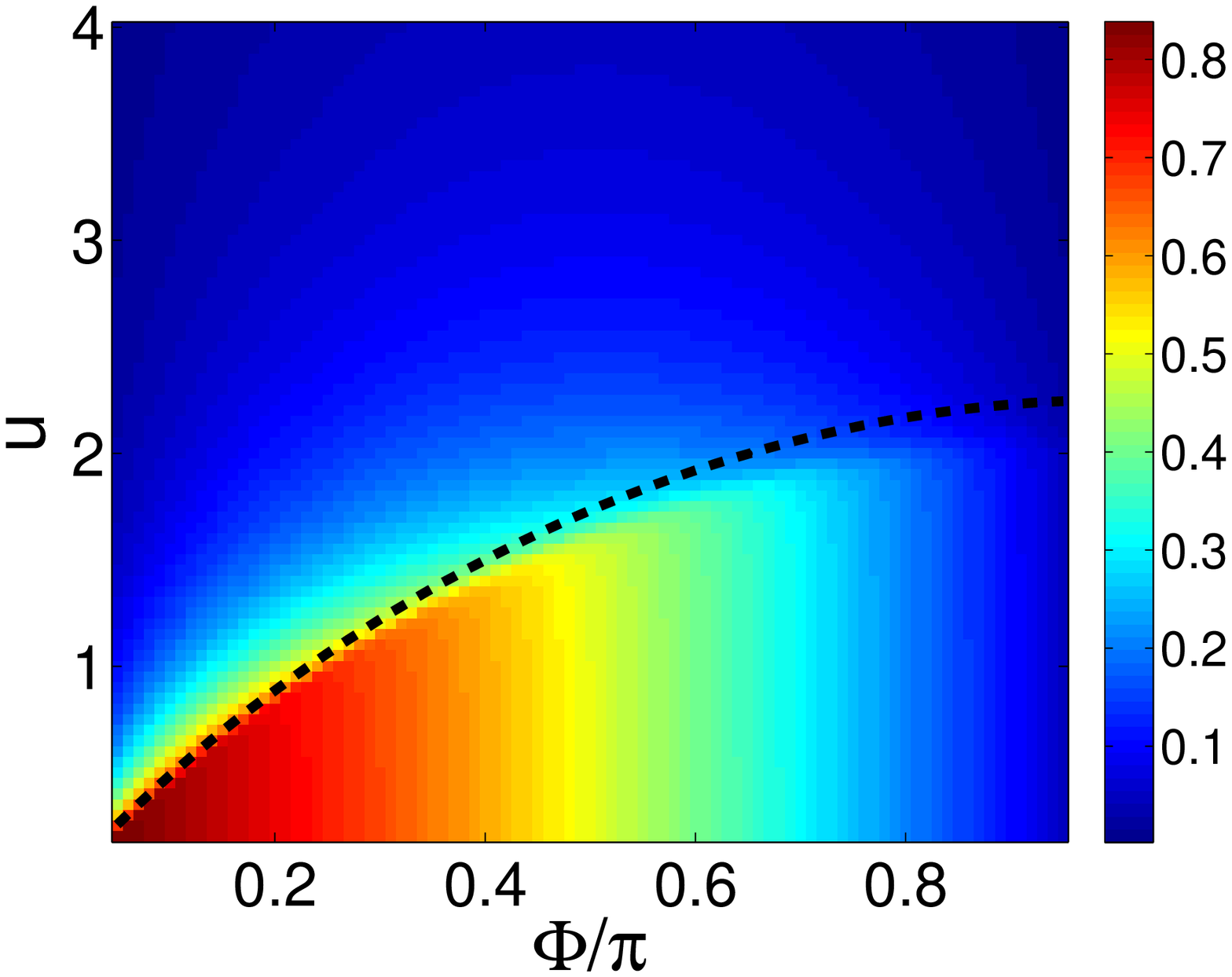}
\includegraphics[width=5cm]{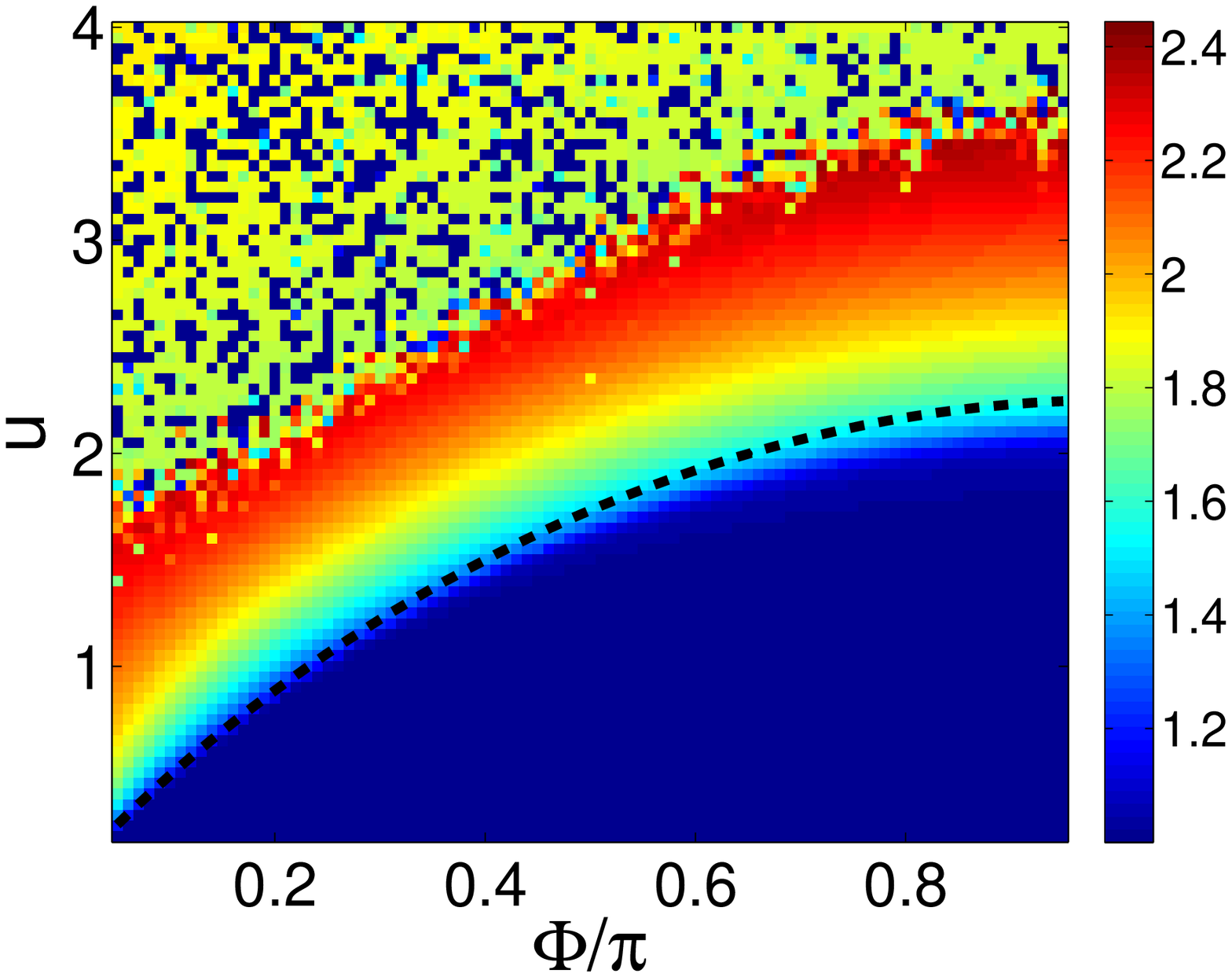}
\includegraphics[width=5cm]{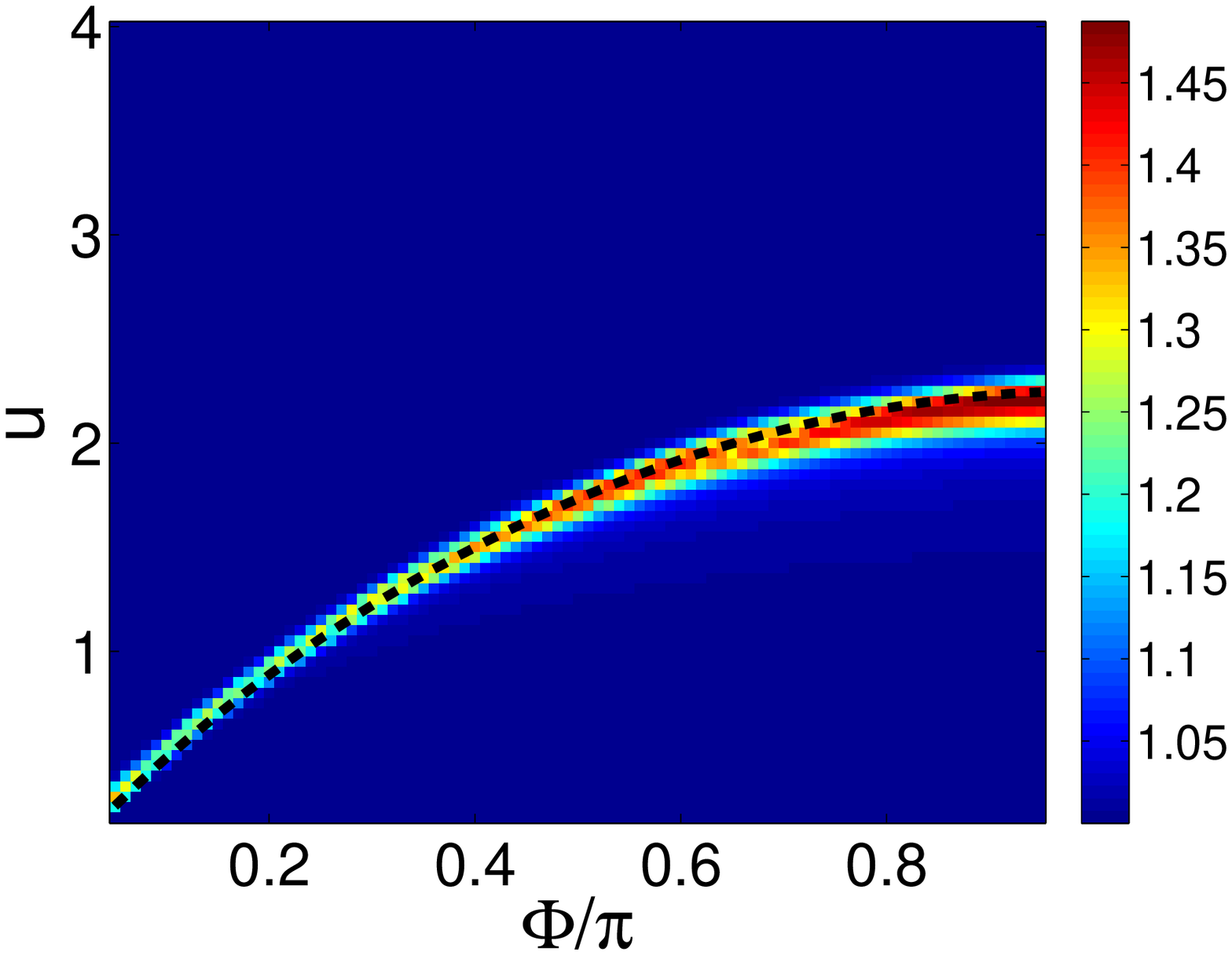}

\ \\

(d) Current $\langle I \rangle_{-}$ 
\hspace{2.5cm} (e) $1/S_{-}$  
\hspace{4cm} (f) correlation
\hspace{0.5cm}

\includegraphics[width=5cm]{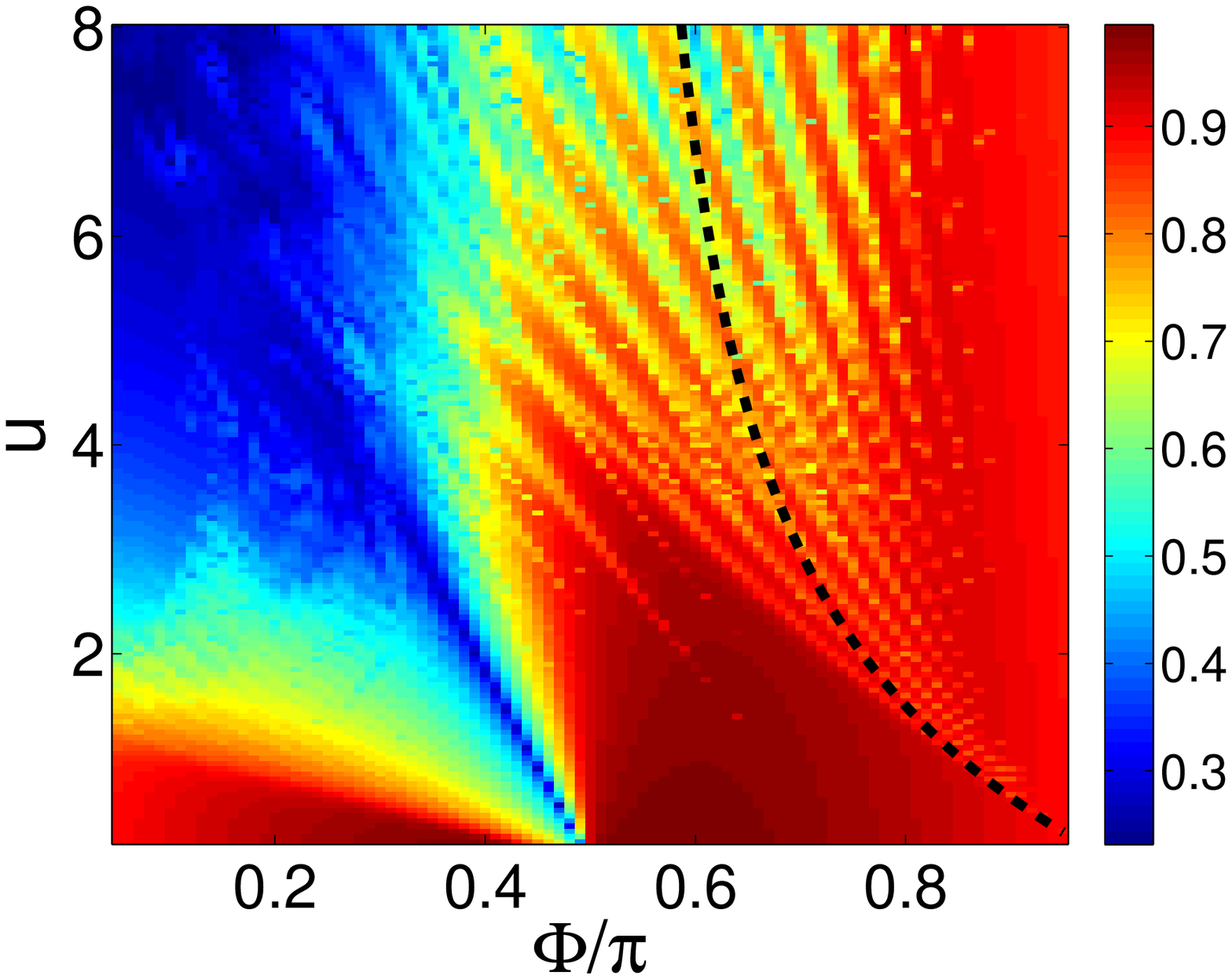}
\includegraphics[width=5cm]{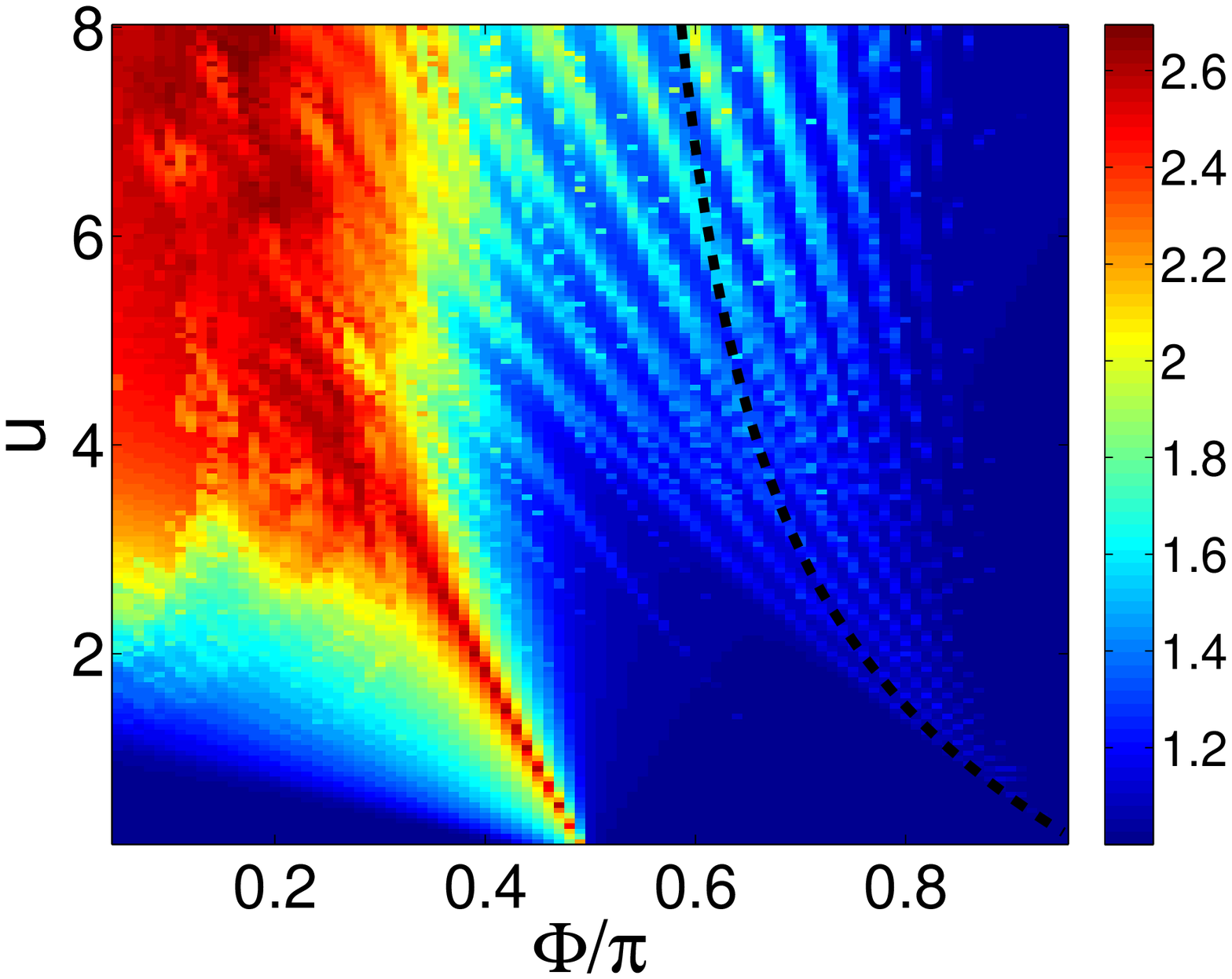}
\includegraphics[width=5cm]{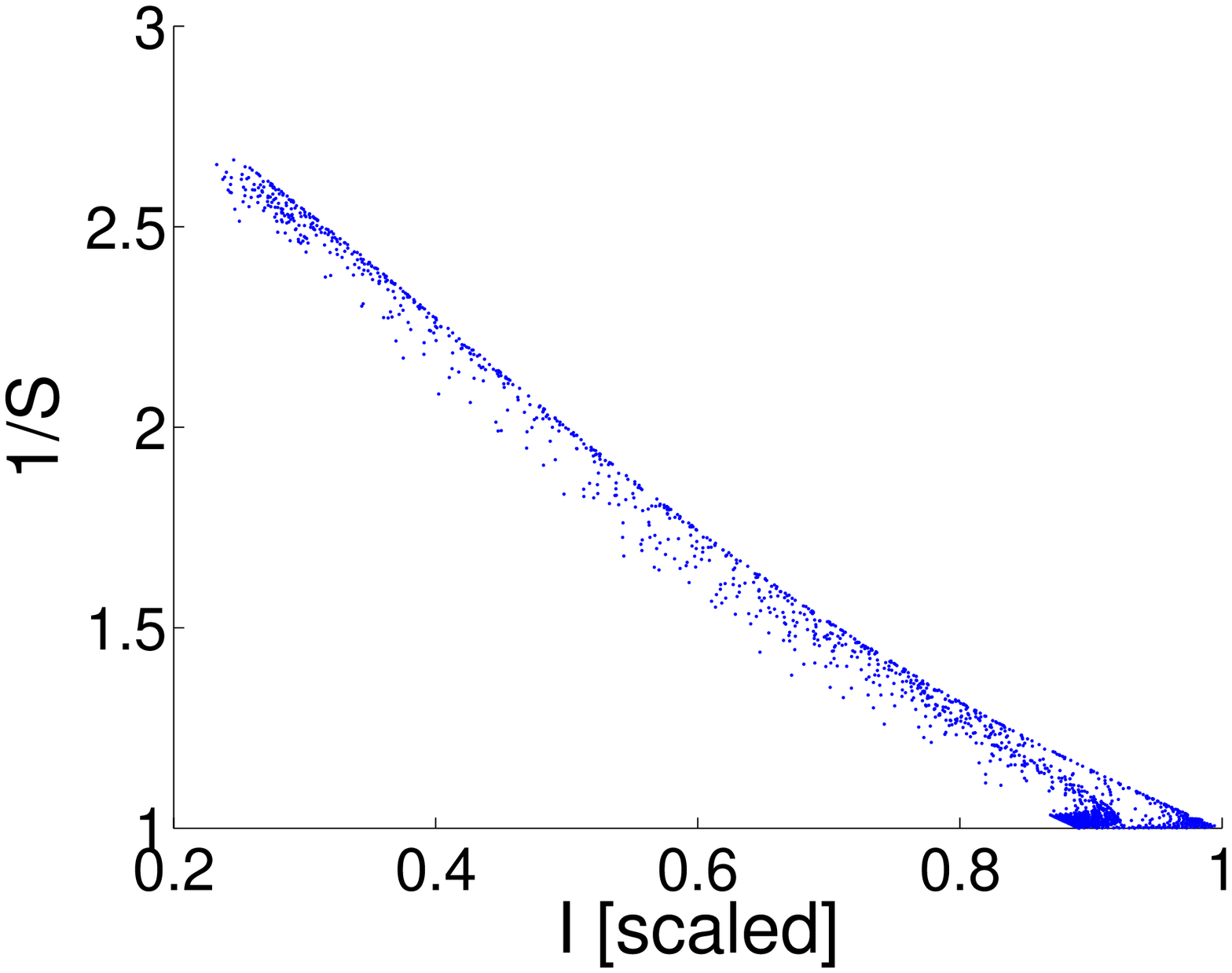}

\caption{(Color online)
We consider a trimer with $N=42$ bososns. 
In (a) The current of the upper-energy state is imaged 
as a function of $(\Phi,u)$. It becomes vanishingly small 
once the self-trapping border (dashed line) is crossed.
In (b) we plot the $1/S$ of the same state, 
while in (c) we repeat the calculation after adding 
some weak disorder {$\epsilon {\sim} 0.01$} (see text).  
Panels (d) and (e) are the same as (a) and (b) 
but for the intermediate state that has 
the maximal current. This is a metastable vortex-state 
if ${S\sim1}$. 
Note that the classical stability border (dashed line)
fails to provide a valid prediction for loss 
of purity, which is strongly correlated with  
low~$I$ values as demonstrated in panel~(f).} 

\label{f5}
\end{figure*}

\section{Self trapping}

{Let us first recall the well-known phase-space analysis of the $M{=}2$ dimer.
A concise detailed account can be found in Section~II of \cite{csd}.}
For ${u<1}$ the energy landscape stretches between a minimum 
that corresponds to condensation in the lower orbital, and a maximum that corresponds to
condensation in the upper orbital.       
For ${u>1}$ the maximum bifurcates and accordingly 
there are two elliptic islands of self-trapped motion.
{See Fig.1 of \cite{csd} for illustration.}

Going back to the trimer, one {realizes} that the dimer 
type bifurcation at ${u=1}$ takes place along 
the edges of the upper energy surface~$V_{+}(r)$.
{This bifurcation is schematically illustrated 
in going from \Fig{f4}ac to \Fig{f4}bd.  
Namely, for ${u>1}$ there are two maxima along  
each edge. But these maxima are merely the corners 
of the central region. The maximum energy  
is located higher: either in the center (\Fig{f4}b)
or bifurcated into three fixed-points (\Fig{f4}d).}  
Hence ${u=1}$ is not the threshold for self-trapping.

Self-trapping in the trimer is related to 
the stability of the $r_0$ fixed-point in the upper 
energy surface. If  ${V_{+}''(r_0)>0}$ 
this upper vortex-state bifurcates into 3 maxima 
that support ``self-trapped" states, 
also known as bright solitons.   
{This bifurcation is schematically illustrated 
in going from \Fig{f4}ab to \Fig{f4}cd.}  
The self-trapping transition is reflected in the 
current dependence as demonstrated in \Fig{f5}.
Namely, the current in panel (a) becomes 
very small once the classical stability border 
({dashed} curve) is crossed. In panel (b) we confirm  
that the loss of stability is reflected by loss 
of purity: once the $r_0$ fixed-point looses  
stability the vortex-state is replaced by 3 soliton-band-states 
that stretch over the 3 fixed-point.

In \Fig{f5}c we repeat the calculation 
as in \Fig{f5}b, but with added 
weak disorder. {Namely, we add small on-site 
energy-shifts $\epsilon_j a_j^{\dag}a_j$ 
to the Hamiltonian \Eq{e1}}
in order to break the translational 
invariance of the system.
These added random shifts are much smaller 
than the inter-site hopping~$K$.  
They do not affect the stability of 
the vortex states but they prevent 
the formation of a soliton band. 
Note that for ${M\gg1}$ the soliton 
band would be exponentially narrow.
Due to the added weak diorder  
the soliton-band states disintegrate 
into self-trapped coherent states 
that have high purity.
In fact some self-trapping also happens
in the numerics of panel~(b) 
due to the finite accuracy of the computer.
{The same numerical issue is well known  
with regard to self-trapping in the 
dimer system \cite{nst1,nst2}.}

The above bifurcation scenario appears contradictory to common wisdom.
For an $M$ site ring  self-trapping is anticipated when 
the self-induced potential is deeper than the binding energy, 
leading to the condition ${u^{cl}>1}$, where 
\beq
{u^{cl} \ \ = \ \ \frac{2}{M} u~.}
\eeq
Contrary to this naive expectation, \Eq{e6} for the trimer  
implies that the threshold for self-localization is vanishingly 
small at the limit ${\Phi\rightarrow 0}$. 
The explanation for this anomaly is as follows: 
for $\Phi=0$ the ${m=\pm1}$ angular-momentum orbitals are degenerate, 
hence any small $U$ results in 3 maxima in $V_{+}(r)$. 
In the case of an $M$~site ring any small $U$ results in $M$ maxima. 
But if $M\gg1$ these maxima represent states 
that have very weak modulation in the site occupation
rather than self-trapping.

\section{Mott transition}

The BEC ground-state corresponds to the minimum of 
the lower $V(r)$ surface, which is an elliptic island.
If $u$ is too large this island becomes too small  
to support a coherent-state and the ground-state 
number-squeezes towards a Fock-basis state. 
For a Bose-Hubbard dimer ($M{=}2$) the ground-state becomes 
a fragmented Fock-state of ${50\%-50\%}$ site 
occupation if ${u>N^2}$. See e.g. \cite{csd}. 
More generally, for an $M$ site ring, 
the Mott superfluid-insulator transition 
is controlled  by the {\em quantum} dimensionless parameter 
\be{3}
{u^{qm} \ \ = \ \ \frac{Mu}{N^2}~.}
\eeq
As $u^{qm} > 1$  the ground-state {looses} its one-body coherence   
and approaches a Fock-state of equal site occupation.

The regimes of \Fig{f4} that are implied by 
the ``classical" stability analysis,  
are related to the topology of phase-space:
they can be resolved if $\hbar$ is reasonably small, 
but do not depend on $\hbar$. 
(Note again that $\hbar=1/N$ reflects the total number 
of particle in the system).
In contrast, the ``quantum" Mott transition 
has to do with having a finite $\hbar$.
{As $u$ is increased the area of the 
lower stability island becomes smaller.}
Due to having a finite Planck-cell, 
the shrinking lower surface of the Hamiltonian 
cannot support a coherent-state 
if $u$ becomes too large. Instead there appear 
a glassy set of low energy fragmented Fock-states. 
The transition is observed in \Fig{f2}. 
{Namely, as implied by the term ``glassiness",  
one observes in panel~(e) a non-zero density 
of low energy states, as opposed to panels (a-d) 
of the same figure where the ground state is 
situated in a well defined location.}
The observed glassiness is due to the possibility 
to play with the occupation whenever $N/M$ 
is not an integer, or due to having some on-site 
disorder.
The Mott transition becomes ``sharp" only 
in the thermodynamic limit of having large~$M$, 
keeping $N/M$ constant.

\section{Metastability}

Having examined the minimum of the lowest Peierls-Nabarro surface $V_0(r)$ 
that undergoes a Mott transition, 
and the maximum of upper surface $V_{+}(r)$ in connection with self-trapping, 
we turn our attention to the intermediate surface $V_{-}(r)$. 
We observe in \Fig{f2} that a dynamically (meta)stable vortex-states 
can be found in the middle of the energy spectrum.
The $\Phi$ threshold for stabilization is deduced 
from the condition ${V_{-}''(r_0)>0}$ provided $V_{-}(r)$ 
is a ``lower" surface. This border is illustrated 
schematically in \Fig{f4} and analytically in \Fig{f5} (dashed black line). 
{Unlike self-trapping, here this (classical) border is barely reflected 
in the numerical results. One observes that large 
current that is supported by high purity vortex-state 
appear well beyond the expected stability region.}
Thus a coherent-solution can be quantum-mechanically 
stabilized in a flat landscape by interference.
We refer to this as ``quasi-stability".
 
What we call quasi-stability is the possibility 
to have a quantum coherent eigenstate 
that is supported by an unstable fixed-point. 
If we have an {\em hyperbolic} fixed-point 
that is immersed in a chaotic sea this is known 
as ``quantum scarring" {\cite{scars1,scars2}}. 
Another well known example for quasi-stability 
is the Anderson strong-localization effect {\cite{AL}.}

Let us point out two examples for quasi-stability  
in the BHH context.
The simplest example is apparently the condensation of bosons 
in the upper orbital of a dimer \cite{csf}: this is formally 
the same as saying that the upper position of a pendulum 
is quasi-stable rather than unstable.  
An additional example is encountered in the case of a kicked dimer \cite{ckt}, 
which is a manifestation of quantum scarring \cite{scars1,scars2}. 
In both examples the quasi-stability is related 
to the low participation number (PN).
The PN characterizes a coherent-state 
that is situated on the hyperbolic point; 
it estimates how many eigenstates 
appear in its spectral decomposition.
In the first example the deterioration of the purity 
is small because PN$\,\sim\log(N)$ rather than PN$\,\sim\sqrt{N}$, 
while in the second example  PN$\,\sim N$ 
with a prefactor that depends on the Lyapunov exponent.

It seems to us that in the present analysis 
the traditional paradigms for quantum quasi-stability do not apply. 
{The natural tendency is to associate the 
observed quasi-stability with quantum scarring,  
and to proceed with the analysis as in \cite{ckt}.
If this were the case, quasi-stability would be 
related to the curvature at $r_0$. This curvature 
becomes worse as we cross $\Phi=\pi/2$. But the 
patterns in \Fig{f5} are not correlated with $V''(r_0)$.} 
We thus conclude that a new paradigm is required.

\section{Concluding remarks}

We presented a comprehensive overview of a minimal model for a superfluid circuit. 
Contrary to the conventional picture we observe that self-trapping 
can occur for arbitrarily small interaction, 
and that unstable vortex-states can become quasi-stable. 
These anomalies reflect the mesoscopic nature of the device: 
effects that are related to orbital-degeneracy 
and quantum-scarring cannot be neglected.

A two orbital approximation as in \cite{Brand2}
does not qualify as a minimal model for a superfluid circuit, 
but it captures one essential ingredient: 
as a parameter is varied a fixed-point can undergo 
a bifurcation. Specifically a vortex-state 
can bifurcate into solitons. In the absence of 
symmetry breaking the bifurcation is into $M$~solitons, 
as illustrated in \Fig{f2} in going from regime~(a) to regime~(c).
These solitons form a band unless the displacement symmetry 
is broken, say by disorder. The 2~orbital approximation 
assumes such symmetry breaking, and provides a simplified 
local description of the bifurcation. 
A global description of phase-space topology 
requires to go beyond the 2~orbital approximation. 
Then one encounters eigenstates that dwell in the 
chaotic sea. Consequently one can regard the trimer 
as a bridge towards the classical and the thermodynamic limits.  
All the required ingredients are here:
the topology and the underlying mixed phase-space.

We note that in a former work \cite{Ghosh} it has been argued
that for $\Omega=0$ metastable vortex-states would be found 
provided ${M>4}$. The argument explicitly assumes ${\Omega=0}$, 
and it is based on a semiclassical (mean field) stability analysis.   
We find that for ${\Omega\ne 0}$ the semiclassical 
stability analysis allows metastable vortex-states
for ${M=3}$ as well. The results of our semiclassical analysis 
are summarized by the ${(\Omega,u)}$ regime diagram of \Fig{f4}.

But when we go to the quantum analysis we find 
that the physical picture is further 
modified quite dramatically due to the manifestation 
of quantum interference effect that is not expected 
on the basis of a mean-field theory. 
This unexpected quasi-stability is effective 
enough to stabilize metastable vortex-states even 
if the device is non-rotating!  \\

{\em Note added in proof.-- } 
Relevant works have been brought to
our attention recently \cite{P1,P2,P3}.

\newpage
{\bf Acknowledgments.-- }
{We thank James Anglin, Tony Leggett, and Parag Ghosh 
for useful communication regrading the observation  
that for $\Omega=0$ metastable vortex-states would 
be found provided ${M>4}$.}  
We thank Igor Tikhonenkov and Christine Khripkov 
for sharing numerical codes 
that had been used in \cite{trm}.
This research was supported by the Israel Science Foundation (grant Nos. 29/11 and 346/11) 
and by the US-Israel Binational Science Foundation (grant No. 2008141).


\end{document}